\documentclass[10pt,a4paper]{article}

\usepackage{graphicx,color}
 \usepackage{bm}
  \usepackage{bbm}
   \usepackage{amsmath}
    \usepackage{amssymb}    
     \usepackage{pifont}
\usepackage{stmaryrd}
\usepackage[export]{adjustbox}
\usepackage{braket}
\usepackage{standalone}
\usepackage{slashed}
\usepackage{multirow}
\usepackage{tabu}
\usepackage{hhline}
\usepackage{colortbl}

\newcommand{\nn}{\nonumber}

\newcommand{\Tr}{\mathrm{Tr}}

\renewcommand{\(}{\left(}
\renewcommand{\)}{\right)}
\renewcommand{\[}{\left[}
\renewcommand{\]}{\right]}

\newcommand{\ta}{\left(}
\newcommand{\tc}{\right)}
\newcommand{\qa}{\left[}
\newcommand{\qc}{\right]}

\usepackage[numbers,sort&compress]{natbib}
\usepackage[colorlinks=true,linkcolor=blue,linktoc=all,citecolor=blue]{hyperref}

\reversemarginpar
\begin{document} 

\vspace{-2.0cm}
\begin{flushright}
  IPARCOS-UCM-23-038
\end{flushright}
\vspace{.1cm}

\begin{center} 

  {\large \bf  Transverse momentum dependent factorization for SIDIS at next-to-leading power}
  \vspace{.7cm}
  \title{}


Simone~Rodini\footnote{\href{mailto:simone.rodini@desy.de}{simone.rodini@desy.de}},

Alexey~Vladimirov\footnote{\href{mailto:alexeyvl@ucm.es}{alexeyvl@ucm.es}},


\vspace{.3cm}
$^1${\it CPHT, CNRS, Ecole Polytechnique, Institut Polytechnique de Paris, Route de Saclay, 91120 Palaiseau, France}\\
$^2${\it Departamento de F\'isica Te\'orica \& IPARCOS, Universidad Complutense de Madrid, E-28040 Madrid, Spain}
\end{center}   

\begin{center}
  {\bf \large Abstract}\\
\end{center}
The semi-inclusive deep-inelastic scattering (SIDIS) is the golden process for investigating the nucleon's transverse momentum-dependent (TMD) structure. We present the complete expression for SIDIS structure functions in the TMD factorization formalism at the next-to-leading power. All perturbative elements of the factorization theorem -- coefficient functions and evolution kernels -- are presented at one-loop accuracy. We found several differences with earlier derivations, which are due to accounting for nontrivial kinematics of the quark-gluon interference terms. As a side result, we present the definition and evolution of twist-three TMD fragmentation functions, including the leading-order evolution kernel.

\newpage

\tableofcontents




\allowdisplaybreaks

\section{Introduction}

The transverse momentum dependent (TMD) factorization theorem emerged from the resummation formalism \cite{Collins:1981uw, Collins:1984kg} and has since evolved into an independent and powerful tool for studying the nucleon's internal structure. While the leading power (LP) TMD factorization theorem \cite{Becher:2010tm, Collins:2011zzd, Echevarria:2011epo} has proven predictive power for various processes, it also suffers from several theoretical inconsistencies and practical limitations (see, f.i., discussions in refs.\cite{Angeles-Martinez:2015sea, Grewal:2020hoc, Scimemi:2019cmh, Bacchetta:2019sam, Bacchetta:2019qkv}). Some of these issues are expected to be resolved or mitigated by including power corrections. For that reason, within the last few years, many groups have concentrated efforts on the problem of TMD factorization at sub-leading power \cite{Balitsky:2017gis, Inglis-Whalen:2021bea, Vladimirov:2021hdn, Balitsky:2021fer, Ebert:2021jhy, Rodini:2022wic, Gamberg:2022lju}. In this work, we continue our earlier development on this subject \cite{Vladimirov:2021hdn, Rodini:2022wic, Rodini:2022wki}, and derive the cross-section for the semi-inclusive deep-inelastic scattering (SIDIS) at next-to-leading power (NLP) TMD factorization and next-to-leading order (NLO) perturbative accuracy.

The foundation of this work is ref.\cite{Vladimirov:2021hdn}, where the general operator expression for the hadronic tensor for Drell-Yan, semi-inclusive deep-inelastic scattering (SIDIS) and semi-inclusive annihilation at next-to-leading power (NLP) and next-to-leading order was derived. The method proposed in ref. \cite{Vladimirov:2021hdn}, named TMD operator expansion, employs the background-field approach and follows the concept of the operator product expansion in its original form \cite{Gross:1973ju}. Being formulated in the position space, it allows an easier treatment of higher twist operators that have involved support properties in the momentum space. The transition from the operator formulation to the cross-section requires several technically involved steps, the procedure for which have been elaborated in \cite{Rodini:2022wic, Rodini:2022wki}. In this work, we compile previous results and obtain the expression for the structure functions of SIDIS at NLP. Let us emphasize that the derivation made in ref.\cite{Vladimirov:2021hdn} is not complete, in the sense that it does not prove the factorization theorem formally. It is assumed that hadron wave-functions could be separated. To justify this assumption one must demonstrate the cancellation of Glauber regions, such as it done for the LP case \cite{Collins:1985ue,Collins:1988ig}.

The primary motivation is to establish the foundation for future phenomenological studies of power corrections in the TMD factorization approach. The more general derivation method used in the present study uncovers several contributions overlooked in previous computations. A notable difference is the contribution of genuine twist-three distributions with vanishing gluon momentum. These terms are crucial to ensure theoretical consistency and may also have numerical significance, although it remains to be seen if they can be observed in modern experiments \cite{Aghasyan:2011ha, COMPASS:2014kcy, HERMES:2020ifk}. The upcoming generation of SIDIS experiments \cite{AbdulKhalek:2021gbh, Anderle:2021wcy} will provide an opportunity to explore these effects in greater detail. Apart from these overlooked terms, the results of the present work independently reproduces the earlier computations, such as \cite{Mulders:1995dh, Boer:2003cm, Bacchetta:2006tn}.

The inclusion of NLO terms requires a strict separation of contributions by their field-theoretical properties. In particular, the conventional approach of distinguishing TMD distributions by their dynamical twist \cite{Mulders:1995dh, Boer:2003cm, Goeke:2005hb, Bacchetta:2006tn, Arnold:2008kf} is inadequate, as it leads to an incomplete basis of operators that becomes evident at NLO, where new types of operators arise. To avoid it, we separate the contributions by the TMD-twist \cite{Vladimirov:2021hdn}. The TMD distributions with distinct TMD twists are independent physical functions. Consequently, the parts of the hadron tensor associated with different TMD twists are self-contained, with non-mixing nonperturbative distributions, independent coefficient functions, and distinct interpretations. The scaling properties of each term are known, which makes possible to consistently account for QCD evolution in the phenomenology of NLP observables.

The paper accumulates all practically important information about the NLP terms. In Sec.~\ref{sec:NLP-gen}, we provide a summary of the findings from \cite{Vladimirov:2021hdn, Rodini:2022wki}, presenting the expressions for the hadronic tensor for SIDIS at both LP and NLP powers and discussing their general properties. In Sec.~\ref{sec:TMDs}, we present the relevant TMD distributions, including their definition, interpretation, and evolution equations. The computation of the SIDIS structure functions is detailed in Sec.~\ref{sec:SIDIS}, leading to the actual results presented in Sec.~\ref{sec:final-expression}. Additional technical information and intermediate results can be found in the appendices. In appendix~\ref{app:complex-structure}, we discuss the derivation of genuine hadronic tensor from the bare results of ref.\cite{Vladimirov:2021hdn}. In Appendix~\ref{app:evolution}, we present expressions for evolution kernels for TMDPDF and TMDFFs of twist-three. Notably, the evolution kernels for TMDFFs presented in Appendix~\ref{app:TMDFF-evol} are original results derived in this paper. Lastly, Appendix~\ref{app:KPC} outlines the contributions to kinematic power corrections from various sources and explains their recombination.

\section{TMD factorization at NLP}
\label{sec:NLP-gen}

In this section, we present the expression for the hadronic tensor at NLP TMD factorization derived in refs.~\cite{Vladimirov:2021hdn, Rodini:2022wki}. We provide an explanation of its structure, but for the details of the derivation we refer the reader to the original papers. The expressions presented in this section are used as the starting point for the following derivation of the cross-section.

The hadronic tensor for SIDIS at NLP TMD factorization is naturally split into three terms
\begin{eqnarray}\label{W:W-all}
W^{\mu\nu}=
\int d\tilde x d\tilde z\, \delta\(\tilde x+\frac{q_+}{P_+}\)\delta\(\tilde z-\frac{p_h^-}{q_-}\)
\int \frac{d^2b}{(2\pi)^2}e^{i(q_T b)}
~\frac{\tilde z}{2}\Big[
\widetilde{W}^{\mu\nu}_{\text{LP}}
+
\widetilde{W}^{\mu\nu}_{\text{kNLP}}
+
\widetilde{W}^{\mu\nu}_{\text{gNLP}}
+...\Big],
\end{eqnarray}
where $q$, $P$, and $p_h$ are momenta of photon, target, and produced hadrons, correspondingly. The labels $\pm$ and $T$ stay for the light-cone and transverse components of vectors. As usual, we denote two light-cone vectors by $n^\mu$ and $\bar n^\mu$ ($n^2=\bar n^2=0$, and $(n\bar n)=1$), and the light-cone projections as $v^+=(nv)$, $v^-=(\bar nv)$, for any vector $v^\mu$. The details of the definition of kinematic variables are unimportant in this section. Therefore, we postpone their precise definition till sec. \ref{sec:SIDIS-kinematics}. 

In eqn.~(\ref{W:W-all}), the term $\widetilde{W}^{\mu\nu}_{\text{LP}}$ represents the LP part. The NLP terms $\widetilde{W}^{\mu\nu}_{\text{kNLP}}$ and $\widetilde{W}^{\mu\nu}_{\text{gNLP}}$  represent the kinematic and genuine power corrections, correspondingly. The dots stay for the higher power corrections. In the following subsections, we present these terms one-by-one providing their explicit expressions and explaining their features and origin.

\subsection{LP term} 
The LP terms has been derived in many works, see f.i., \cite{Boer:2003cm, Collins:2011zzd, Echevarria:2011epo, Becher:2010tm}. It can be written in the following compact form
\begin{eqnarray}\label{def:W-LP}
\widetilde{W}_{\text{LP}}^{\mu\nu}
=
|C_1(\mu^2,Q^2)|^2
\sum_{n,m}\Big[
\Tr\(\gamma^\mu \overline{\Gamma}_m^+\gamma^\nu \overline{\Gamma}_n^-\)
\Phi_{11}^{[\Gamma^+_n]}(x,b;\mu,\zeta) 
\Delta_{11}^{[\Gamma_m^-]}(z,b;\mu,\bar \zeta)
\\\nn 
+
\Tr\(\gamma^\mu \overline{\Gamma}_n^-\gamma^\nu \overline{\Gamma}_m^+\)
\overline{\Phi}_{11}^{[\Gamma^+_n]}(x,b;\mu,\zeta)
\overline{\Delta}_{11}^{[\Gamma_m^-]}(z,b;\mu,\bar \zeta)
\Big].
\end{eqnarray}
Here, $\Phi_{11}$ ($\overline{\Phi}_{11}$) and $\Delta_{11}$ ($\overline{\Delta}_{11}$) are the TMDPDF and TMDFF of twist-two for (anti-)quarks. Their field-theoretical definition is provided in sec.\ref{sec:TMDs}. The variable $Q^2$ is the hard scale of the process, associated with the photon's invariant mass, $q^2=-Q^2$. The scale $\mu$ is the factorization scale for the separation of hard and collinear modes. The scales $\zeta$ and $\bar \zeta$ are factorization scale of collinear and anti-collinear modes (rapidity separation) \cite{Echevarria:2012js, Chiu:2012ir}. These scales must satisfy $\zeta \bar \zeta=Q^4$. The factor $|C_1|^2$ is the LP hard coefficient function. The NLO expression for $|C_1|^2$ in the SIDIS case reads (the expression for $C_1$ is given in eqn.(\ref{def:C1}))
\begin{eqnarray}\label{def:C1^2}
|C_1(\mu^2,Q^2)|^2=1+2a_s(\mu)C_F\(-\ln^2\(\frac{Q^2}{\mu^2}\)+3\ln\(\frac{Q^2}{\mu^2}\)-8+\frac{\pi^2}{6}\)+\mathcal{O}(a_s^2),
\end{eqnarray}
where $C_F=(N_c^2-1)/2N_c$ (for $SU(N_c)$ gauge group), $a_s=g^2/(4\pi)^2$ is the QCD coupling constant. Nowadays, the expression for $C_1$ is known at N$^4$LO \cite{Lee:2022nhh}.

The matrices $\Gamma^\pm$ in eqn.~(\ref{def:W-LP}) stay for the full set of Dirac gamma-matrices projecting good components of quark spinors. The quark TMD distributions are given by matrix elements of quark-anti-quark fields whose spinor indices are contracted as $\Phi^{[\Gamma]}\sim \langle \bar q \Gamma q\rangle$, see eqn.~(\ref{def:PHI11}). In this work we discuss only twist-two and twist-three TMD distributions. These distribution incorporate only the good-components of the spinors (with respect to vector $n$). The space of $\Gamma$-matrices that survive in this contraction, is four-dimensional. The standard basis in this space is denoted by $\Gamma^+$, and the elements of this basis are
\begin{eqnarray}\label{def:Gamma+}
\Gamma^+= \{\gamma^+, \gamma^+\gamma^5, i\sigma^{\alpha+}\gamma^5\},
\end{eqnarray}
where $\alpha$ is a transverse index. The decomposition of any matrix $A$ in this space reads
\begin{eqnarray}
A=\frac{1}{2}\sum_{n} \overline{\Gamma}^-_n A^{[\Gamma^+_n]},\qquad 
A^{[\Gamma]}=\frac{1}{2}\Tr\(\Gamma A\),
\end{eqnarray}
where
\begin{eqnarray}\label{def:GammaOver-}
\overline{\Gamma}^-= \{\gamma^-,-\gamma^-\gamma^5, -i\sigma^{\alpha-}\gamma^5\},
\end{eqnarray}
and index $n$ runs though all elements (\ref{def:Gamma+}) and (\ref{def:GammaOver-}) are in one-to-one correspondence. Conventionally, TMDPDF are constructed with collinear, and thus projected by $\Gamma^+$ (\ref{def:Gamma+}). The TMDFFs are constructed with anti-collinear fields and thus their components projected with respect to vector $\bar n$ by $\Gamma^-$. The matrices $\Gamma^-$ and $\overline{\Gamma}^+$ are obtained from (\ref{def:Gamma+}, \ref{def:GammaOver-}) replacing $n\leftrightarrow \bar n$.

\subsection{Kinematic NLP term} 
The kinematic part of NLP term reads
\begin{eqnarray}\label{def:Wk}
\widetilde{\mathcal{W}}^{\mu\nu}_{\text{kNLP}}(y)
&=&-i|C_1(\mu^2,Q^2)|^2\sum_{n,m}\Bigg\{
\\\nn &&
\frac{\bar n^\mu \Tr[\gamma^{\rho}\overline\Gamma_m^+ \gamma^\nu \overline\Gamma_n^-]+\bar n^\nu \Tr[\gamma^\mu \overline\Gamma_m^+\gamma^\rho \overline\Gamma_n^-]}{q_-}
\Phi_{11}^{[\Gamma^+_n]}
\(\frac{\partial}{\partial b^\rho}-\frac{\partial_\rho \mathcal{D}}{2}\ln\(\frac{\zeta}{\bar \zeta}\)\)
\Delta_{11}^{[\Gamma_m^-]}
\\\nn &&
+
\frac{\bar n^\mu \Tr[\gamma^{\rho}\overline\Gamma_n^- \gamma^\nu\overline\Gamma^+_m]
+\bar n^\nu \Tr[\gamma^\mu \overline\Gamma_n^-\gamma^\rho \overline\Gamma_m^+]}{q_-}
\overline{\Phi}_{\bar n11}^{[\Gamma_n^+]}
\(\frac{\partial}{\partial b^\rho}-\frac{\partial_\rho \mathcal{D}}{2}\ln\(\frac{\zeta}{\bar \zeta}\)\)\overline{\Delta}_{n11}^{[\Gamma_m^-]}
\\\nn &&
+
\frac{n^\mu \Tr[\gamma^{\rho}\overline\Gamma_m^+ \gamma^\nu\overline\Gamma^-_n]+n^\nu \Tr[\gamma^\mu \overline\Gamma_m^+\gamma^\rho \overline\Gamma_n^-]}{q_+}
\Delta_{n11}^{[\Gamma_m^-]}
\(\frac{\partial}{\partial b^\rho}+\frac{\partial_\rho \mathcal{D}}{2}\ln\(\frac{\zeta}{\bar \zeta}\)\)\Phi_{\bar n11}^{[\Gamma_n^+]}
\\\nn &&
+
\frac{n^\mu \Tr[\gamma^{\rho}\overline\Gamma_n^- \gamma^\nu \overline\Gamma_m^+]
+
n^\nu \Tr[\gamma^\mu \overline\Gamma_n^-\gamma^\rho \overline\Gamma_m^+]}{q_+}
\overline{\Delta}_{n11}^{[\Gamma^-_m]}
\(\frac{\partial}{\partial b^\rho}+\frac{\partial_\rho \mathcal{D}}{2}\ln\(\frac{\zeta}{\bar \zeta}\)\)
\overline{\Phi}_{\bar n11}^{[\Gamma_n^+]}
\Bigg\},
\end{eqnarray}
where we have omit the arguments $(\tilde x,b;\mu,\zeta)$ for all TMDPDFs, and $(\tilde z,b;\mu,\bar\zeta)$ for all TMDFFs for brevity. The derivative acts on the following term, and
\begin{eqnarray}\label{dD}
\partial_\rho \mathcal{D}=\frac{\partial}{\partial b^\rho} \mathcal{D}(b,\mu)=2b_\rho \frac{\partial }{\partial b^2}\mathcal{D}(b,\mu),
\end{eqnarray}
with $\mathcal{D}$ being the Collins-Soper kernel. The Collins-Soper kernel is the nonperturbative function that describes the evolution of TMD distributions, see eqn.(\ref{evol:rapidity}). The second equality in eqn.~(\ref{dD}) is due to the fact that the Collins-Soper kernel depends only on $b^2$.

The terms propotional to $\partial_\rho \mathcal{D}$ are the remnants of cancellation of the special rapidity divergences, which appear in the (bare) genuine NLP term. These terms play special role in the TMD factorization -- they restore the broken boost invariance at NLP \cite{Rodini:2022wic}. Thanks to these terms, the expression (\ref{def:Wk}) is independent on the choice of $\zeta$ (while the $\bar \zeta$ is fixed due to $\zeta\bar \zeta=Q^4$). It can be checked explicitly by rescaling $\zeta\to \alpha \zeta$ and $\bar \zeta\to \bar \zeta/\alpha$, and differentiating the result by $\alpha$. Note, that the point $\zeta=\bar \zeta=Q^2$, which is commonly used for the phenomenology, these terms vanish.

The kinematic power correction has the same nonperturbative content and the hard coefficient function as the LP term. It is not accidental but the requirement of the electro-magnetic gauge invariance (charge conservation) for the hadronic tensor, which requires
\begin{eqnarray}\label{qW=0}
q^{\mu}W_{\mu\nu}=W_{\nu\mu}q^{\mu}=0.
\end{eqnarray}
For the hadronic tensor in the $b$-space this equation reads
\begin{eqnarray}\label{qW(b)=0}
\hat q^{\mu}\widetilde{W}_{\mu\nu}=\(q^+ \bar n^{\mu}+q^- n^\mu+i\frac{\partial}{\partial b_\mu}\)\widetilde{W}_{\mu\nu}=0.
\end{eqnarray}
It is straightforward to check that the LP term (\ref{def:W-LP}) does not satisfy (\ref{qW(b)=0}), since
\begin{eqnarray}\label{EM:1}
\hat q_{\mu}\widetilde{W}^{\mu\nu}_{\text{LP}}=
i\frac{\partial}{\partial b^{\mu}}\widetilde{W}^{\mu\nu}_{\text{LP}}\neq 0.
\end{eqnarray}
In the momentum space, the expression (\ref{EM:1}) is proportional to $q_T$. Thus, the LP term of TMD factorization violates the electro-magnetic gauge invariance, but the violation is next-to-leading in the power counting. Taking into account the kinematic NLP term, the gauge invariance is improved, 
\begin{eqnarray}\label{EM:2}
\hat q_{\mu}\(\widetilde{W}^{\mu\nu}_{\text{LP}}+\widetilde{W}^{\mu\nu}_{\text{kNLP}}\)=
i\frac{\partial}{\partial b^{\mu}}\widetilde{W}^{\mu\nu}_{\text{kNLP}}\neq 0.
\end{eqnarray}
Here, the violation is of N$^2$LP order (or $\sim q_T^2/Q$ in the momentum space). The accounting of kinematic N$^2$LP term would improve charge conservation up to N$^3$LP, etc. Complete restoration of electro-magnetic gauge invariance requires an infinite chain of kinematic power corrections. The restoration of the gauge invariance also guaranties that the coefficient function equals to LP coefficient functions at all perturbative orders. This has been checked explicitly in ref.\cite{Vladimirov:2021hdn} at NLO.

\subsection{Genuine NLP term}

The genuine NLP term contains the novel nonperturbative distributions -- TMDPDFs and TMDFFs of twist-three. These are quark-gluon-quark correlators, that depend on three momentum-fraction variables (with their sum equal to zero). The TMDPDF of twist-three are discussed in details in the dedicated work \cite{Rodini:2022wki}. The synopsis of that study and its extension to TMDFFs is presented in the sec.\ref{sec:TMDs}. 

The hadronic tensor for genuine NLP term derived in ref.\cite{Vladimirov:2021hdn} could not be used as it is presented, because it contains complex-valued functions, and distributions with indefinite parity. Therefore, prior to operation it should be presented in the form with explicit complex structure. The details of this transformation are presented in the appendix \ref{app:complex-structure}. The resulting expression is
\begin{eqnarray}\label{def:WgNLP}
\widetilde{W}_{\text{gNLP}}&=&i \sum_{n,m}
\Bigg\{
\int_{-\infty}^\infty d\hat u_1d\hat u_2d\hat u_3\,
\delta(\hat u_1+\hat u_2+\hat u_3) \delta(\tilde x-\hat u_3)\Bigg[
\\\nn && 
T^{\mu\nu\rho}_-(\bar n,n)\(
\mathbb{C}_R(x,\hat u_2) \mathbf{\Phi}_{\rho,\oplus}^{[\Gamma^+_n]} \Delta_{11}^{[\Gamma_m^-]}
+\pi\mathbb{C}_I(x,\hat u_2) \mathbf{\Phi}_{\rho,\ominus}^{[\Gamma^+_n]}\Delta_{11}^{[\Gamma_m^-]}\)
\\\nn && 
+iT^{\mu\nu\rho}_+(\bar n,n) \(
\pi \mathbb{C}_I(x,\hat u_2) \mathbf{\Phi}_{\rho,\oplus}^{[\Gamma^+_n]}\Delta_{11}^{[\Gamma_m^-]}
-\mathbb{C}_R(x,\hat u_2) \mathbf{\Phi}_{\rho,\ominus}^{[\Gamma^+_n]}\Delta_{11}^{[\Gamma_m^-]}\)
\\\nn && 
+T^{\mu\nu\rho}_-(n,\bar n)\(
\mathbb{C}_R(x,\hat u_2)\overline{\mathbf{\Phi}}_{\rho,\oplus}^{[\Gamma^+_n]}\overline{\Delta}_{11}^{[\Gamma^-_m]}
+\pi\mathbb{C}_I(x,\hat u_2)\overline{\mathbf{\Phi}}_{\rho,\ominus}^{[\Gamma^+_n]}\overline{\Delta}_{11}^{[\Gamma^-_m]}\)
\\\nn && 
+iT^{\mu\nu\rho}_+(n,\bar n)\(
\pi\mathbb{C}_I(x,\hat u_2)\overline{\mathbf{\Phi}}_{\rho,\oplus}^{[\Gamma^+_n]}\overline{\Delta}_{11}^{[\Gamma^-_m]}
-\mathbb{C}_R(x,\hat u_2)\overline{\mathbf{\Phi}}_{\rho,\ominus}^{[\Gamma^+_n]}\overline{\Delta}_{11}^{[\Gamma^-_m]}\)
\Bigg]
\\\nn &&
+ \int_{-\infty}^\infty \frac{d\hat w_1d\hat w_2d\hat w_3}{|\hat w_1|}\,\delta\(\frac{1}{\hat w_1}+\frac{1}{\hat w_2}+\frac{1}{\hat w_3}\)
\delta(\tilde z-\hat w_3)\Bigg[
\\\nn &&
T^{\mu\nu\rho}_-(\bar n,n) \mathbb{C}_2(z,\hat w_2) \Phi_{11}^{[\Gamma^+_n]}\mathbf{\Delta}_{\rho,\oplus}^{[\Gamma_m^-]}
-iT^{\mu\nu\rho}_+(\bar n,n)
\mathbb{C}_2(z,\hat w_2) \Phi_{11}^{[\Gamma^+_n]}\mathbf{\Delta}_{\rho,\ominus}^{[\Gamma_m^-]}
\\\nn &&
+
T^{\mu\nu\rho}_-(n,\bar n)
\mathbb{C}_2(z,\hat w_2) \overline{\Phi}_{11}^{[\Gamma^+_n]}\overline{\mathbf{\Delta}}_{\rho,\oplus}^{[\Gamma_m^-]}
-iT^{\mu\nu\rho}_+(n,\bar n)
\mathbb{C}_2(z,\hat w_2) \overline{\Phi}_{11}^{[\Gamma^+_n]}\overline{\mathbf{\Delta}}_{\rho,\ominus}^{[\Gamma_m^-]}
\Bigg]\Bigg\}.
\end{eqnarray}
where $\mathbf{\Delta}_{\rho,\oplus}$ and $\mathbf{\Delta}_{\rho,\ominus}$ are TMDFFs of twist-three, $\mathbf{\Phi}_{\rho,\oplus}$ and $\mathbf{\Phi}_{\rho,\ominus}$ are TMDPDFs of twist-three. All these distributions are the quark-gluon-quark correlators. The formal definition of these distributions is given in sec.\ref{sec:TMD-formal}. The tensor $T_\pm^{\mu\nu\rho}$ are
\begin{eqnarray}
T^{\mu\nu \rho}_\pm(\bar n,n)=\(\frac{\bar n^\mu }{q_-}-\frac{n^\mu}{q^+}\)\Tr[\gamma^{\rho}\overline{\Gamma}_m^+ \gamma^\nu \overline{\Gamma}_n^-]
\pm\(\frac{\bar n^\nu}{q_-}-\frac{n^\nu}{q_+}\)\Tr[\gamma^\mu \overline{\Gamma}_m^+\gamma^\rho \overline{\Gamma}_n^-].
\end{eqnarray} 
For simplicity of presentation we omit the arguments of TMD distributions. They are: $(\hat u_1,\hat u_2,\hat u_3,b;\mu,\zeta)$ for all twist-three TMDPDFs,
$(\hat w_1,\hat w_2,\hat w_3,b;\mu,\bar \zeta)$ for all twist-three TMDFFs,
$(\tilde x,b;\mu, \zeta)$ for all twist-two TMDPDFs, and $(\tilde z,b;\mu,\bar \zeta)$ for all twist-two TMDFFs. The variables $\hat u$ and $\hat w$ are the momentum-fractions of partons. In particular, the variables $\hat u_2$ and $\hat w_2$ are the momentum-fractions of gluon field. The delta-functions represent the conservation of momentum, they can be used to eliminate any two integrations in a convenient manner. We do it after the determination of support ranges for TMDFFs in sec.\ref{sec:interpretationTMDFF}.

The coefficient functions are
\begin{eqnarray}\label{CR}
&&\mathbb{C}_{R}(x,u_2)=
\frac{1}{(u_2)_+}+a_s\Bigg\{
2\frac{C_F}{(u_2)_+}\(-\mathbf{L}^2+2\mathbf{L}-\frac{11}{2}+\frac{\pi^2}{6}\)
\\\nn &&
\qquad
+2\(C_F-\frac{C_A}{2}\)\frac{1}{(u_2)_+}\frac{x}{u_2}\[\(\mathbf{L}-2+\frac{1}{2}\ln\(\frac{|x+u_2|}{x}\)\)\ln\(\frac{|x+u_2|}{x}\)
-\frac{\pi^2}{2}\theta(-x-u_2)\]
\\\nn &&
\qquad
+C_A\frac{x}{x+u_2}\[-\(\frac{\ln|u_2|}{u_2}\)_++\frac{\ln x}{(u_2)_+}
+\frac{\pi^2}{2}\delta(u_2)\]
\Bigg\}+\mathcal{O}(a_s^2)
\\\label{CI}
&&
\mathbb{C}_{I}(x,u_2)=
\delta(u_2)+a_s\Bigg\{
2C_F\[\delta(u_2)\(-\mathbf{L}^2+2\mathbf{L}-\frac{15}{2}+\frac{\pi^2}{6}\)\]
\\\nn &&
\qquad
+2\(C_F-\frac{C_A}{2}\)\[\delta(u_2)\mathbf{L}
+\frac{1}{(u_2)_+}\frac{x}{u_2}\(
\theta(-x-u_2)(\mathbf{L}-2)+\theta(-x-u_2)\ln\(\frac{|x+u_2|}{x}\)\)\]
\\\nn &&
\qquad
+C_A\[\delta(u_2)(\ln x+2)-\frac{\theta(u_2)}{(u_2)_+}\frac{x}{x+u_2}\]\Bigg\}
+\mathcal{O}(a_s^2),
\\\label{C2}
&&\mathbb{C}_2(z,w_2)=
1+a_s\Big[
2C_F\(-\mathbf{L}^2+2\mathbf{L}-\frac{11}{2}+\frac{\pi^2}{6}\)
\\\nn &&\qquad
+2\(C_F-\frac{C_A}{2}\)\frac{w_2}{z}\(\mathbf{L}-2+\frac{1}{2}\ln\(1-\frac{z}{|w_2|}\)\)\ln\(1-\frac{z}{|w_2|}\)
\\\nn &&\qquad
-C_A\frac{|w_2|}{|w_2|-z}\ln\(\frac{z}{|w_2|}\)\Big]+\mathcal{O}(a_s^2),
\end{eqnarray}
where $C_A=N_c$,  $\mathbf{L}=\ln(Q^2/\mu^2)$, $\theta(x)$ is the Heaviside theta function and ``plus''-distribution is defined as
\begin{eqnarray}
\int du \(f(u)\)_+g(u)=
\int du f(u)\(g(u)-g(0)\).
\end{eqnarray}
These coefficient functions are originated from a single complex-values coefficient function given in eqn. (\ref{app:C2}). The derivation of these expressions from ref.\cite{Vladimirov:2021hdn} is a somewhat non-trivial task and we present it in appendix.\ref{app:complex-structure}. The coefficient functions $\mathbb{C}_R$, $\mathbb{C}_I$, and $\mathbb{C}_2$ cannot be obtained from each other despite stemming from a common origin.

Let us also stress the non-trivial contribution coming from the imaginary parts to the hadronic tensor. The $\mathbb{C}_I$ coefficient function starts at LO with $\delta(\hat u_2)$, i.e. it produces terms with vanishing gluon collinear momenum. This is similar to the famous Qiu-Sterman contribution to the single-spin-asymmetry \cite{Qiu:1991pp}. These contribution were overlooked in all earlier studies of SIDIS at NLP (see f.i. \cite{Bacchetta:2006tn, Boer:2003cm, Ebert:2021jhy}). Note that at NLO similar contribution appears also in the coefficient function $\mathbb{C}_R$. This is expected and it can be traced to the structure of the NLP evolution equations.
The logarithm structure of coefficient functions match the evolution equations for twist-three and twist-two TMD distributions. To confirm it one should differentiate (\ref{def:WgNLP}), use the evolution equations collected in appendix \ref{app:evolution}, and integrate over $\hat u$'s and $\hat w$'s where it is possible. The details of this computation can be found in sec. 9 of ref.\cite{Vladimirov:2021hdn}. 

The genuine NLP term is transverse up to NNLP,
\begin{eqnarray}\label{EM:3}
\hat q_{\mu} \widetilde{W}_{\text{gNLP}}=i\frac{\partial}{\partial b^\mu}\widetilde{W}_{\text{gNLP}}=\mathcal{O}(\text{NNLP}).
\end{eqnarray}
At NNLP there are two independent kinematic power correction parts. One contains only TMD distributions of twist-two and is responsible for the restoration of QED gauge invariance, see eq.(\ref{EM:2}). The second kinematic power correction contains a TMD distribution of twist-two and a TMD distribution of twist-three and restores the gauge invariance of $\widetilde{W}_{\text{gNLP}}$, see eq.(\ref{EM:3}). In this sense each genuine N$^k$LP term is the first term in the series of kinematic power corrections. The coefficient function of such series are the same for all terms (but different for each series).

\section{TMD distributions of sub-leading power}
\label{sec:TMDs}

The description of SIDIS requires two types of correlators: TMDPDF correlators and TMDFF correlators. Their operator definitions are similar, yet they exhibit fundamental differences in their properties and evolution. The NLP cross-section incorporates TMD distributions of twist-two and twist-three. The guiding principle for such decomposition is the scaling properties of distributions. If any two distributions do not mix under the evolution they are entirely independent nonperturbative functions. In our formulation we use the notion of TMD-twist \cite{Vladimirov:2021hdn} to decompose the operator into independent parts. So, the twist-two distributions $\Phi_{11}$ are independent objects compared to twist-three distributions $\mathbf{\Phi}_{\oplus}$ and $\mathbf{\Phi}_{\ominus}$. 

In the twist-two case, the TMD distributions are simply the correlators of quark fields (multiplied by semi-infinite Wilson lines), so $\Phi_{11}\sim \langle P|\bar q.. q|P\rangle$. There are eight TMDPDFs and two TMDFFs (for spinless particle) \cite{Boer:2003cm, Goeke:2005hb}. In the twist-three case, the operator has an additional insertion of the gluon field, $\Phi^\mu_{\oplus}\sim \langle P|\bar q..F^{\mu+}.. q|P\rangle$. The gluon field can be inserted at either light-cone operators, which lead to plethora of distributions with various properties. In total there are 32 TMDPDFs (for spin-1/2 particle) and 8 TMDFFs (for spinless particle). Luckily, only some of them appear in the SIDIS at NLP.  The TMDPDFs of twist-three and their properties are discussed in a great details in a dedicated work \cite{Rodini:2022wki}. On contrary, the TMDFFs of twist-three are practically unstudied. 

\subsection{Formal definition of twist-three distributions}
\label{sec:TMD-formal}

The guiding principle for the computation of power corrections is the twist-decomposition, which separates the distributions into the independent nonperturbative functions. In expressions (\ref{def:Wk}, \ref{def:WgNLP}) this separation is done using the geometric definition of the TMD-twist. The (geometric) TMD-twist is a conserving QCD quantum number, and thus TMD distributions of different TMD-twists are strictly independent nonperturbative functions. The set of TMD distributions with definite TMD-twist is complete, in the sense that no other distributions could mix with it, and any factorization theorem (of corresponding power) could be described solely in their terms. The theory of TMD-twist decomposition was developed in refs.\cite{Vladimirov:2021hdn, Rodini:2022wki}. The formal definition of twist-three TMD distributions is rather involved, and in fact, unnecessary for a practice. Nonetheless in this section, we present main details of the formal definition, for completeness. For an extended discussion, we refer the interested reader to original works.

All TMD matrix elements have a common general structure. They are correlators of two light-cone operators separated by the transverse distance. This is a consequence counting rules that lead to the TMD factorization. The light-cone operators are semi-compact, in the sense that they contain semi-infinite Wilson lines that continue to the light-cone infinity and has an open color index. The light-cone operators a can be ordered with respect to geometrical twist, which is the ``dimension-minus-spin'' of the operator. The TMD-twist is the pair of numbers (N,M), where N and M are the twists of each semi-compact operators. We use the term ``twist-three'' as a general indication of TMD-twist-(1,2) or TMD-twist-(2,1) when the distinction is inessential.

At NLP only the following semi-compact operators appear
\begin{eqnarray}
\label{definition_SemiCompactOperators}
U_{1,i}(\lambda;b) &=& \mathbf{W}_{\pm}(b)[\pm n\infty+b,\lambda n+b]q_i(\lambda n+b), \\
U^\mu_{2,i}(\lambda_2,\lambda_1;b) &=& \mathbf{W}_{\pm}(b)[\pm n\infty+b,\lambda_2n+b] gF^{\mu+}(\lambda_2n+b)[\lambda_2n+b,\lambda_1n+b]q_i(\lambda_1n+b),\nn\\
\overline{U}_{1,i}(\lambda;b) &=& \bar{q}_i(\lambda n+b)[\lambda n+b,\pm n\infty+b]\mathbf{W}^\dagger_{\pm}(b),\nn\\
\overline{U}^\mu_{2,i}(\lambda_1,\lambda_2;b) &=& \bar{q}_i(\lambda_1n+b)[\lambda_1n+b,\lambda_2n+b]gF^{\mu+}(\lambda_2n+b)[\lambda_2n+b,\pm n\infty+b]\mathbf{W}^\dagger_{\pm}(b), \nn
\end{eqnarray}
where $\lambda$'s are coordinates along the light-cone directions, $b$ is the transverse vector ($b^+=b^-=0$), $\pm$ is selected depending on the process, $F_{\mu\nu}$ is the gluon-field-strength tensor, and $i$ is the spinor index of the quark-field. In the above definitions, the gauge link is defined along the straight path 
\begin{equation}
[a,b] = \mathcal{P}\exp\(  -ig \int_0^1 dt (b^\mu-a^\mu) A_\mu(bt+a(1-t)) \) 
\end{equation}
$\mathbf{W}$ is the transverse gauge link that is required to restore the QCD gauge invariance \cite{Belitsky:2002sm, Idilbi:2010im}
\begin{equation}
\mathbf{W}_{\pm}(b) = \mathcal{P}\exp\(  -ig \int_\infty^1 dt b^\mu A_\mu(bt\pm n\infty ) \).
\end{equation}
The subscripts 1 and 2 of the semi-compact operators refer to the twist of these operators. Importantly,  only the ``good'' spinor components of the operators (\ref{definition_SemiCompactOperators}) have twist 1. These components can be projected by $\gamma^-\gamma^+/2$. This projection reduce the space of possible Dirac combinations to the basis (\ref{def:Gamma+}). The bad spinor components do not have definite geometrical twist and could be simplified using QCD equations of motions. The operators (\ref{definition_SemiCompactOperators}) are defined along the light-cone direction $n$, but the same definition is valid for any direction.

In the case of SIDIS the Wilson lines within the definition of TMDPDFs are pointing to the $+\infty n$, while it is $-\infty n$ for Drell-Yan process. The Wilson lines of TMDFFs are pointing to the $-\infty n$, while it is $+\infty n$ for the annihilation process. 

The TMDPDF of TMD-twist-(1,1) is defined as
\begin{eqnarray}\label{def:PHI11}
\Phi_{11}^{[\Gamma]}(x,b)=\int_{-\infty}^\infty \frac{d\lambda}{2\pi} e^{-ix \lambda P^+} 
\langle P,s|\overline{U}_1(\lambda;b)\frac{\Gamma}{2}U_1(0;0)|P,s\rangle,
\end{eqnarray}
where $|P,s\rangle$ is the hadron state with large component of momentum $P^+$. The twist-three TMDPDFs can have TMD-twist-(1,2) or (2,1), and they are defined as
\begin{eqnarray}\label{def:PHI21}
\Phi_{\mu,21}^{[\Gamma]}(x_1,x_2,x_3,b)&=&\int_{-\infty}^\infty \frac{d\lambda_1 d\lambda_2}{(2\pi)^2} e^{i(x_1\lambda_1+x_2\lambda_2)P^+} 
\langle P,s|\overline{U}_{\mu,2}(\lambda_1,\lambda_2;b)\frac{\Gamma}{2}U_1(0;0)|P,s\rangle,
\\\label{def:PHI12}
\Phi_{\mu,12}^{[\Gamma]}(x_1,x_2,x_3,b)&=&\int_{-\infty}^\infty \frac{d\lambda_1 d\lambda_2}{(2\pi)^2} e^{i(x_1\lambda_1+x_2\lambda_2)P^+} 
\langle P,s|\overline{U}_{1}(\lambda_1;b)\frac{\Gamma}{2}U_{\mu,2}(\lambda_2,0;0)|P,s\rangle,
\end{eqnarray}
where $x_3=-x_2-x_1$. The three-variable notation is a bit redundant, but it is convenient for discussion of properties of TMD distributions. Generally, $x_1$, $x_2$ and $x_3$ can be though as the collinear momentum-fractions of anti-quark, gluon, and quark correspondingly. However, since $x_i$ do not have definite sign, this interpretation is correct only for a specific sign combination. We review the interpretation in sec.\ref{sec:interpretationTMDPDF}.

The distributions $\overline{\Phi}$ are associated with the anti-quark distributions and have inverted order of quark fields
\begin{eqnarray}\label{def:PHI11-bar}
\overline{\Phi}_{11}^{[\Gamma]}(x,b)&=&\int_{-\infty}^\infty \frac{d\lambda}{2\pi} e^{-ix \lambda P^+} 
\Tr\langle P,s|\frac{\Gamma}{2}U_1(\lambda;b)\overline{U}_1(0;0)|P,s\rangle,
\\
\label{def:PHI21-bar}
\overline{\Phi}_{\mu,21}^{[\Gamma]}(x_1,x_2,x_3,b)&=&\int_{-\infty}^\infty \frac{d\lambda_1 d\lambda_2}{(2\pi)^2} e^{i(x_1\lambda_1+x_2\lambda_2)P^+} 
\Tr\langle P,s|\frac{\Gamma}{2} U_{\mu,2}(\lambda_1,\lambda_2;b)\overline{U}_1(0;0)|P,s\rangle,
\\\label{def:PHI12-bar}
\overline{\Phi}_{\mu,12}^{[\Gamma]}(x_1,x_2,x_3,b)&=&\int_{-\infty}^\infty \frac{d\lambda_1 d\lambda_2}{(2\pi)^2} e^{i(x_1\lambda_1+x_2\lambda_2)P^+} 
\Tr\langle P,s|\frac{\Gamma}{2} U_{1}(\lambda_1;b)\overline{U}_{\mu,2}(\lambda_2,0;0)|P,s\rangle,
\end{eqnarray}
where $\Tr$ contracts spinor and color indices. Notice that for all TMDPDFs, the products of operators could be collected into common T-ordering \cite{Jaffe:1983hp}. It gives us the relations
\begin{eqnarray}\nn
\overline{\Phi}^{[\Gamma]}_{11}(x,b)&=&-\Phi^{[\Gamma]}_{11}(-x,-b),
\\\label{phi(-x)=barphi}
\overline{\Phi}^{[\Gamma]}_{12}(x_1,x_2,x_3,b)&=&-\Phi^{[\Gamma]}_{21}(x_3,x_2,x_1,-b),
\\\nn
\overline{\Phi}^{[\Gamma]}_{21}(x_1,x_2,x_3,b)&=&-\Phi^{[\Gamma]}_{12}(x_3,x_2,x_1,-b).
\end{eqnarray}

The TMDFFs are defined analogously to TMDPDFs, but with the hadron moved to a final state. The twist-two TMDFF is
\begin{eqnarray}
\Delta_{11}^{[\Gamma]}(z,b)&=&
\frac{1}{2zN_c}\int_{-\infty}^\infty \frac{d\lambda}{2\pi} e^{i\frac{\lambda p^+}{z}}
\sum_X \Tr\langle 0|\frac{\Gamma}{2}U_1(\lambda;b)|p,X\rangle \langle p,X|\overline{U}_1(0;0)|0\rangle,
\end{eqnarray}
where $1/(2zN_c)$ is a conventional prefactor \cite{Metz:2016swz}. The twist-three TMDFFs are
\begin{eqnarray}\label{def:DELTA21}
\Delta_{21}^{[\Gamma]}(z_1,z_2,z_3,b)&=&
\frac{1}{2|z_1z_2z_3|N_c}\int_{-\infty}^\infty \frac{d\lambda_1 d\lambda_2}{(2\pi)^2} e^{-i\(\frac{\lambda_1}{z_1}+\frac{\lambda_2}{z_2}\)p^+}
\\\nn &&\qquad\qquad
\sum_X \Tr \langle 0|\frac{\Gamma}{2}U_2(\lambda_1,\lambda_2;b)|p,X\rangle \langle p,X|\overline{U}_1(0;0)|0\rangle,
\\
\Delta_{12}^{[\Gamma]}(z_1,z_2,z_3,b)&=&
\frac{1}{2|z_1z_2z_3|N_c}\int_{-\infty}^\infty \frac{d\lambda_1 d\lambda_2}{(2\pi)^2} e^{-i\(\frac{\lambda_1}{z_1}+\frac{\lambda_2}{z_2}\)p^+}
\\\nn &&\qquad\qquad
\sum_X \Tr \langle 0|\frac{\Gamma}{2}U_1(\lambda_1;b)|p,X\rangle \langle p,X|\overline{U}_2(\lambda_2,0;0)|0\rangle,
\end{eqnarray}
where
\begin{eqnarray}
\frac{1}{z_1}+\frac{1}{z_2}+\frac{1}{z_3}=0.
\end{eqnarray}
The factor $1/(2|z_1z_2z_3|N_c)$ mimics the $1/(2zN_c)$ factor for the twist-two distributions, and it is assigned such that the cross-section for LP and genuine NLP have the same common factors. The presence of absolute values is needed since at least one of the $z_i$ is negative.

The barred TMDFFs are associated with anti-quark fragmentation and are defined by exchanging $U\leftrightarrow \overline{U}$. One has
\begin{eqnarray}
\overline{\Delta}_{11}^{[\Gamma]}(z,b)&=&
\frac{1}{2zN_c}\int_{-\infty}^\infty \frac{d\lambda}{2\pi} e^{i\frac{\lambda p^+}{z}}
\sum_X \langle 0|\overline{U}_1(\lambda;b)\frac{\Gamma}{2}|p,X\rangle \langle p,X|U_1(0;0)|0\rangle,
\\
\overline{\Delta}_{21}^{[\Gamma]}(z_1,z_2,z_3,b)&=&
\frac{1}{2|z_1z_2z_3|N_c}\int_{-\infty}^\infty \frac{d\lambda_1 d\lambda_2}{(2\pi)^2} e^{-i\(\frac{\lambda_1}{z_1}+\frac{\lambda_2}{z_2}\)p^+}
\\\nn &&\qquad\qquad
\sum_X \langle 0|\frac{\Gamma}{2}\overline{U}_2(\lambda_1,\lambda_2;b)|p,X\rangle \langle p,X|U_1(0;0)|0\rangle,
\\
\overline{\Delta}_{12}^{[\Gamma]}(z_1,z_2,z_3,b)&=&
\frac{1}{2|z_1z_2z_3|N_c}\int_{-\infty}^\infty \frac{d\lambda_1 d\lambda_2}{(2\pi)^2} e^{-i\(\frac{\lambda_1}{z_1}+\frac{\lambda_2}{z_2}\)p^+}
\\\nn &&\qquad\qquad
\sum_X \langle 0|\frac{\Gamma}{2}\overline{U}_1(\lambda_1;b)|p,X\rangle \langle p,X|U_2(\lambda_2,0;0)|0\rangle.
\end{eqnarray}
Unlike the TMDPDF case, the TMDFF operator cannot be combined into a single T-ordered operator because of the hadron in the final state. For this reason, there is no formal relation similar to (\ref{phi(-x)=barphi}) between $\overline{\Delta}$ and $\Delta$. 

The NLP factorization theorem is formulated in terms of (renormalized) distributions of twist-three is well-defined and finite expression. However, individually the terms of the factorization theorem could diverge in the regime of the gluon momentum-fraction $x_2$ (or $1/z_2$) approaching zero. This is a rapidity divergence (according to the general definition of rapidity divergences \cite{Vladimirov:2017ksc}), but of a different nature in comparison to usual rapidity divergences of TMD factorization. For this reason, it is named ``special'' rapidity divergence. To eliminate the special rapidity divergences, and to make each term of the factorized expression finite, one should add  and subtract certain asymptotic terms. This lead to the \textit{physical} TMD distributions of twist-three. We define \cite{Rodini:2022wki}
\begin{eqnarray}\label{def:subtraction}
\mathbf{\Phi}_{\mu,12}^{[\Gamma]}(x_1,x_2,x_3,b)=
\Phi_{\mu,12}^{[\Gamma]}(x_1,x_2,x_3,b)-[\mathcal{R}_{12}\otimes \Phi_{11}]^{[\Gamma]}_\mu(x_1,x_2,x_3,b),
\\\nn
\mathbf{\Phi}_{\mu,21}^{[\Gamma]}(x_1,x_2,x_3,b)=
\Phi_{\mu,21}^{[\Gamma]}(x_1,x_2,x_3,b)-[\mathcal{R}_{21}\otimes \Phi_{11}]^{[\Gamma]}_\mu(x_1,x_2,x_3,b),
\end{eqnarray}
where $\mathcal{R}$ are kernels that could be computed perturbatively, and $\otimes$ is an integral convolution. These kernels involves the derivative of the Collins-Soper kernel. The LO expressions for kernels $\mathcal{R}$ are given in ref.\cite{Rodini:2022wki}. In the following, we do not emphasize the difference between $\Phi$ and $\mathbf{\Phi}$, using everywhere the usual font, but assuming the physical distributions (with subtraction terms).

The matrix elements (\ref{def:PHI11}-\ref{def:PHI12-bar}) are the results of the TMD-twist decomposition. Each of (\ref{def:PHI11}-\ref{def:PHI12-bar}) is renormalized independently, i.e. the different distributions  do not mix during the evolution process. However, these distributions are impractical. They are not the distributions which represent the physical process. In particular, matrix elements $\Phi_{12}$ and $\Phi_{21}$ have indefinite complexity, and indefinite T-conjugation parity. Under these tranformations they turn to each other with a rather complicated rules, for example $[\Phi_{12}^{[\Gamma]}(x_1,x_2,x_3,b)]^*=\Phi_{21}^{[\gamma^0\Gamma \gamma^0]}(-x_3,-x_2,-x_1,-b)$. The full list of rules is given in ref.\cite{Rodini:2022wki}. For that reason, it is practical (and physically motivated, see sec.\ref{sec:interpretationTMDPDF}) to introduce the following combinations
\begin{eqnarray}\label{def:PHI-plus}
\Phi_{\mu,\oplus}^{[\Gamma]}(x_1,x_2,x_3;b)&=&\frac{\Phi_{\mu,21}^{[\Gamma]}(x_1,x_2,x_3;b)+\Phi_{\mu,12}^{[\Gamma]}(-x_3,-x_2,-x_1;b)}{2},
\\\label{def:PHI-minus}
\Phi_{\mu,\ominus}^{[\Gamma]}(x_1,x_2,x_3;b)&=&i\frac{\Phi_{\mu,21}^{[\Gamma]}(x_1,x_2,x_3;b)-\Phi_{\mu,12}^{[\Gamma]}(-x_3,-x_2,-x_1;b)}{2},
\end{eqnarray}
and the same combinations for $\overline{\Phi}$, $\Delta$ and $\overline{\Delta}$. Under the discrete transformations, $\Phi_{\oplus}$ and $\Phi_{\ominus}$ transform onto them-self. And therefore, their components (\ref{def:TMDsPDF:2:g+}-\ref{def:TMDsPDF:2:s+}) are real-valued and have definite T-parity, and could be used as well-defined TMD distributions. They have somewhat clearer interpretation, and the physical process cross-section is naturally written in the terms of $\Phi_{\oplus}$ and $\Phi_{\ominus}$. The drawback is that $\Phi_{\oplus}$ and $\Phi_{\ominus}$ do not have definite TMD-twist, and thus do not evolve independently. Their evolution equation has a matrix form (\ref{def:evol-PDF}), and $\Phi_{\oplus}$ mixes with $\Phi_{\ominus}$. Still the twist-three distributions do not mix with twist-two distributions, which allows an unambiguous separation of kinematic and genuine parts.

We emphasize that the formal definitions presented in this sections are important for the derivation and justification of the factorization theorem. However, for any practical applications one does need to take into account all these details. Practically, the only important information is the parametrization and the evolution equations that are presented in the following sections.

\subsection{Interpretation of TMDPDFs}
\label{sec:interpretationTMDPDF}

The TMDPDFs of twist-two are usually associated with the distribution of partons within the hadron with a given momentum-fraction $x$ and transverse momentum $k_T$. To recover this interpretation one firsts chooses a physical gauge in which the staple-like Wilson lines becomes unity (light-cone gauge with appropriate boundary conditions). One then inserts the full set of states in-between quark fields and observe that
\begin{eqnarray}\label{interp}
\Phi_{11}(x,k_T)\propto \int d^2b d\lambda  e^{ix \lambda P^++i(bk_T)}\langle P|\bar q(0)|X\rangle\langle X|q(\lambda n+b)|P\rangle,
\end{eqnarray}
where $\Phi_{11}(x,k_T)$ is the Fourier image of $\Phi_{11}(x,b)$. The particles in the final state must have positive energy, which restricts $1-x>0$. The similar analyses for $\overline{\Phi}_{11}$ gives $1+x>0$, and together with (\ref{phi(-x)=barphi}) results into $-1<x<1$. The partonic process for $\overline{\Phi}_{11}$ is the same as for (\ref{interp}) but with exchanged quark and anti-quark fields.

In this way, the TMDPDF is the squared amplitude of a process $h(P)\to q(xP+k_T)+X$, i.e. are the distributions of partons with momentum $xP+k_T$ within the hadron. It can be drawn as
\begin{align}
\Phi_{11}(x>0,k_T) &\propto \includegraphics[valign=c]{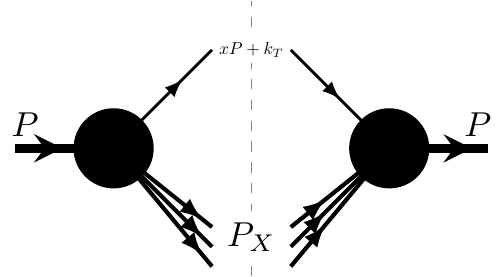}.
\end{align}
Correspondingly, the anti-quark distribution $\overline{\Phi}$ depicts the process
\begin{align}
\Phi_{11}(x<0,k_T) &\propto \includegraphics[valign=c]{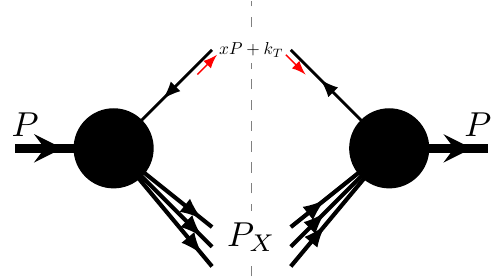}.
\end{align}
So, the sign of $x$ gives interpretation of the underling process.

A very similar consideration provides the support properties and interpretation for the TMDPDFs of twist-three. The variables $x_i$ are subject to constraint
\begin{eqnarray}\label{TMDPDF-support}
x_1+x_2+x_3=0,\qquad -1<x_1,x_2,x_3<1.
\end{eqnarray}
Note, that it implies that at least one of $x$'s is negative. The twist-three distributions describe interference processes, in which, e.g., a quark-gluon pair is emitted and only a quark is re-absorbed. For that reason they could not be interpreted as parton densities, but are probability amplitudes. Similarly to the twist-two case the interpretation depends on the signs of the momentum fractions. Furthermore, transverse momenta of partons associated with operator $U_2$ or $\overline{U}_2$ are integrated such that $k_1+k_2=k_T$.

It is important to mention that the points $x_i=0$ are regular points of the TMDPDFs, which can non-vanishing at these points. The evolution equations (\ref{def:evol-PDF}) are discontinuous at these points. Nonetheless, it is expected that the subtraction term (\ref{def:subtraction}) makes the TMDPDFs of twist-three continuous. At the moment, it is not clear how it is realised practically.

Each combination of signs $x_i \lessgtr 0$ (there are six such combinations) results into a separate underlying parton process \cite{Jaffe:1983hp, Rodini:2022wki}, for $\Phi_{12}$ and $\Phi_{21}$. Here are several examples
\begin{align}
\Phi_{21}(x_1<0,x_2<0,x_3>0,k_T) &\propto \includegraphics[valign=c]{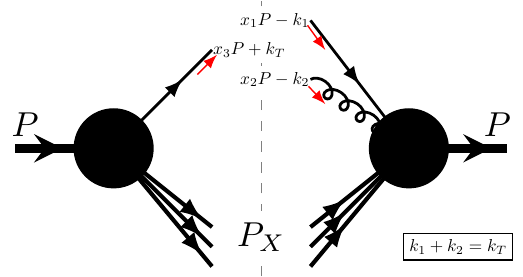}, \\
\Phi_{12}(x_1>0,x_2>0,x_3<0,k_T) &\propto \includegraphics[valign=c]{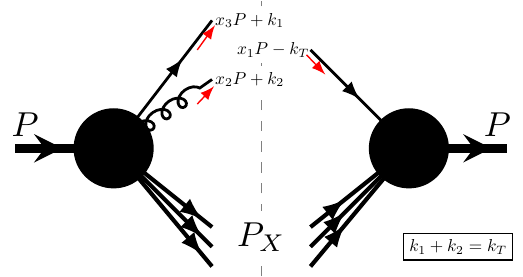},\\
\Phi_{21}(x_1>0,x_2<0,x_3>0,k_T) &\propto \includegraphics[valign=c]{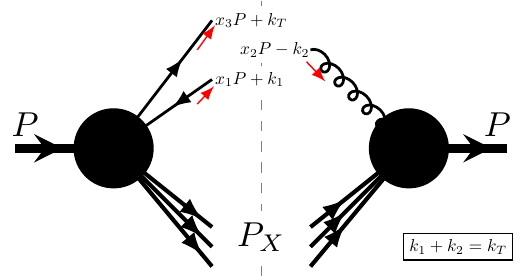}.
\end{align}
Since the twist-three PDF are interference processes they do not have any ``good'' properties, and are complex-values functions. 

On contrary the combinations $\oplus$ and $\ominus$ are real-valued functions, and one can expect a reasonable interpretation for them. Indeed, combining the interpretations for $\Phi_{12}$ and $\Phi_{21}$, we find that $\Phi_{\oplus}$ and $\Phi_{\ominus}$ are the real and imaginary part (respectively) of the interference processes. For instance
\begin{align}
\Phi_\oplus(x_1<0,x_2<0,x_3>0,k_T) &\propto \text{Re} \ta \includegraphics[valign=c]{Figures/PDFtw3_parton_interpret1.pdf}\tc,\\
\Phi_\ominus(x_1<0,x_2<0,x_3>0,k_T) &\propto \text{Im} \ta \includegraphics[valign=c]{Figures/PDFtw3_parton_interpret1.pdf}\tc,
\end{align}
and similar for other combinations of signs. This interpretation picture also holds for anti-quark distributions $\overline{\Phi}_{\oplus}$ and $\overline{\Phi}_{\ominus}$ but with exchange of quark and anti-quark fields.

\begin{figure}[t]
\centering
\includegraphics[width=0.4\textwidth]{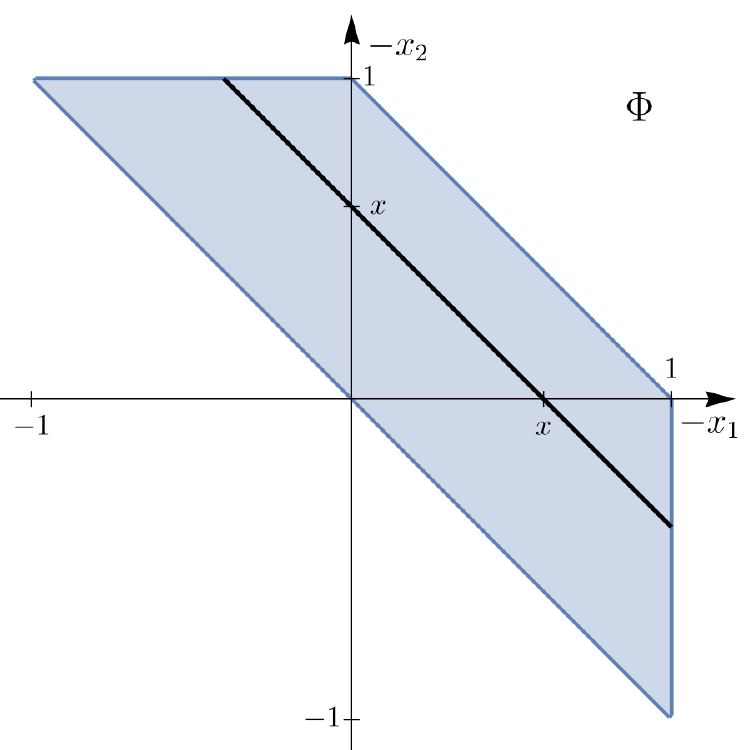}
~~~~~
\includegraphics[width=0.4\textwidth]{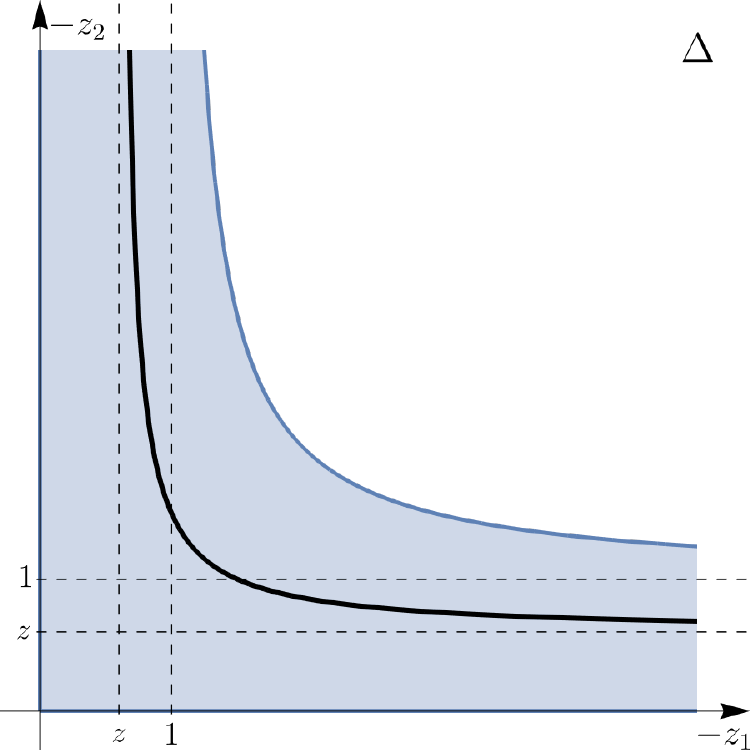}
\caption{Support regions for TMDPDF $\Phi_\bullet(x_1,x_2,x)$ of twist-three (left) and TMDFF $\Delta_\bullet(z_1,z_2,z)$ of twist-three (right), $x>0$ and $z>0$. The $\bullet$ is $\oplus$, $\ominus$ or $21$. The solid black line shows the of constant $x=-x_1-x_2$ and $z=\frac{z_1z_2}{z_1+z_2}$. The integral convolutions that appear at NLP (in coefficient functions and in the evolution kernels) evaluates the distributions along these lines. In the case of TMDFF the region expands to infinities with hyperbolas approaching dashed lines asymptotically.}
\label{fig:support}
\end{figure}

Generally, the distributions $\Phi_{\oplus}$ and $\Phi_{\ominus}$ are defined for all six sign-combinations of $x_i$'s (\ref{TMDPDF-support}). However, the kinematics of the SIDIS (and also Drell-Yan) fixes $x_3=x>0$. Therefore, there are only three active sectors $(x_1<0,x_2<0)$, $(x_1>0,x_2<0)$, and $(x_1<0,x_2>0)$. The shape of this domain is shown in fig.\ref{fig:support}(left). All integrations that are present in the factorization theorem (such as convolution with the coefficient functions, or with the evolution kernel) preserve the value of $x_3=x$, but involves all available sectors of $(x_1,x_2)$. The trajectory of these integrations is shown in fig.\ref{fig:support} by the black line.

\subsection{Interpretation of TMDFFs}
\label{sec:interpretationTMDFF}

The consideration of TMDFFs follows the same general pattern as for TMDPDF case. There is however an important difference. Namely, there is no connection between quark and anti-quark distributions, because corresponding fields are separated by the produced-hadron state and could not migrate to different sides of the cut. For that reason, the kinematics of partonic sub-process restricts $0<z<1$ for both $\Delta_{11}$ and $\overline{\Delta}_{11}$. These partonic processes can be depicted as
\begin{align}
\Delta_{11}(z,k_T) &\propto \includegraphics[valign=c]{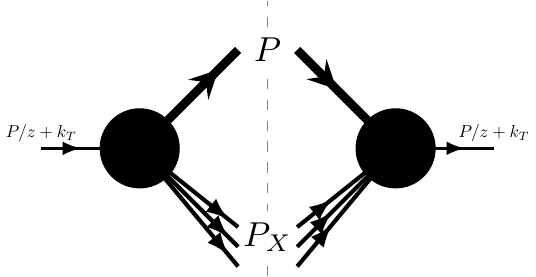},
\\
\overline{\Delta}_{11}(z,k_T) &\propto \includegraphics[valign=c]{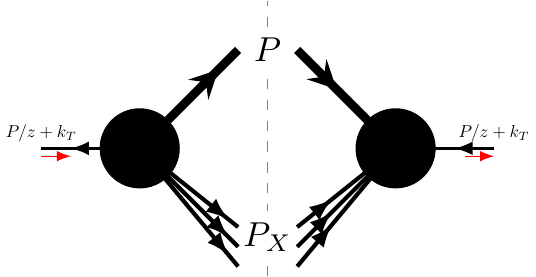}.
\end{align}
I.e. they could be interpreted as the ``probability'' of a parton with momentum $P/z+k_T$ to fragment into the hadron with momentum $P$.

In contrast to TMDPDFs, the TMDFFs of twist-three do not posses any symmetry in their variables and they are non-vanishing only for a single sign-combination:
\begin{eqnarray}\label{support:Delta12}
\Delta_{12}(z_1,z_2,z_3,b)\neq 0, \quad &\text{for}&~-1<z_1<0,~~z_{2,3}>0\text{~~and~~} \frac{1}{z_1}+\frac{1}{z_2}+\frac{1}{z_3}=0,
\\\nn
\Delta_{21}(z_1,z_2,z_3,b)\neq 0, \quad &\text{for}&~0<z_3<1,~~z_{1,2}<0\text{~~and~~} \frac{1}{z_1}+\frac{1}{z_2}+\frac{1}{z_3}=0.
\end{eqnarray}
The support regions for $\overline{\Delta}_{12}$ and $\overline{\Delta}_{21}$ are analogous. Combining equations (\ref{support:Delta12}) and (\ref{def:PHI-plus}, \ref{def:PHI-minus}), we find that that the support regions for $\Delta_\oplus$ is
\begin{eqnarray}\label{support:Delta}
\Delta_{\oplus}(z_1,z_2,z_3,b)\neq 0, \quad \text{for}~0<z_3<1,~~z_{1,2}<0\text{~~and~~} \frac{1}{z_1}+\frac{1}{z_2}+\frac{1}{z_3}=0.
\end{eqnarray}
Furthermore, the support regions for $\Delta_\ominus$, $\overline{\Delta}_\oplus$, and $\overline{\Delta}_\ominus$ coincides with (\ref{support:Delta}). Note, that the kinematics of SIDIS fixes $z_3=z>0$ (similarly to the TMDPDF case), and the value of $z_3$ is preserved by all integrations in at NLP. The values of $z_{1,2}$ are restricted as $z_{1,2}<-z$. The domain of definition is shown in fig.\ref{fig:support}(right), with the integration trajectory shown by black line.

The inequalities in eqn.~(\ref{support:Delta}) are strict. It implies that
\begin{eqnarray}\label{support:TMDFF=0}
\Delta_{12}(z_1,z_2,0)=\Delta_{12}(z_1,0,z_3)=\Delta_{21}(0,z_2,z_3)=\Delta_{21}(z_1,0,z_3)=0.
\end{eqnarray}
This is linked to the fact that at these points the particles in the intermediate state $|X\rangle$ must be all massless, on the mass-shell and carry no traverse momenta \cite{Meissner:2008yf}. 

The fact that $z$'s have fixed signs does not leave a lot of possibilities for the interpretation. The $\Delta_\oplus$ and $\Delta_\ominus$ distributions are the real and imaginary part of the interference amplitudes
\begin{align}
\Delta_{\oplus}(z_1,z_2,z_3,k_T) &\propto \text{Re}\ta\includegraphics[valign=c]{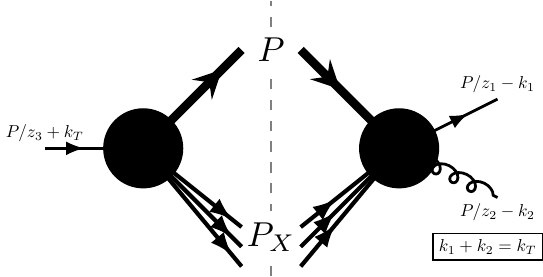}\tc, \\
\Delta_{\ominus}(z_1,z_2,z_3,k_T) &\propto\text{Im}\ta\includegraphics[valign=c]{Figures/FFtw3_parton_interpret1.pdf}\tc,
\end{align}
and similar (with exchange of quark and anti-quark) for $\overline{\Delta}$'s. In contrast to TMDPDF all other combinations of processes are forbidden.

\subsection{Parametrization of TMD distributions}

The parameterization of twist-two TMD correlator was established long ago \cite{Mulders:1995dh, Boer:1997nt}. In the position space it reads \cite{Boer:2011xd, Scimemi:2018mmi}
\begin{eqnarray}
\label{def:TMDPDF11:g+}
\Phi^{[\gamma^+]}_{11}(x,b)&=&f_1(x,b)+i\epsilon^{\mu\nu}_T b_\mu s_{T\nu}M f_{1T}^\perp(x,b),
\\\label{def:TMDPDF11:g+5}
\Phi^{[\gamma^+\gamma^5]}_{11}(x,b)&=&\lambda g_{1}(x,b)+i(b \cdot s_T)M g^\perp_{1T}(x,b),
\\\label{def:TMDPDF11:s+}
\Phi^{[i\sigma^{\alpha+}\gamma^5]}_{11}(x,b)&=&s_T^\alpha h_{1}(x,b)-i\lambda b^\alpha M h_{1L}^\perp(x,b)
\\\nn && +i\epsilon^{\alpha\mu}b_\mu M h_1^\perp(x,b)-\frac{M^2 b^2}{2}\(\frac{g_T^{\alpha\mu}}{2}-\frac{b^\alpha b^\mu}{b^2}\)s_{T\mu}h_{1T}^\perp(x,b),
\end{eqnarray}
where $b^2<0$, and
\begin{eqnarray}\label{def:gT}
g_T^{\mu\nu}=g^{\mu\nu}-n^\mu \bar n^\nu-\bar n^\mu n^\mu,\qquad\epsilon_T^{\mu\nu}=\epsilon^{\mu\nu-+}    ,
\end{eqnarray}
$M$ is the mass of hadron, and $s_T$ and $\lambda$ are components of the hadron's spin-vector
\begin{eqnarray}\label{def:sT}
S^\mu=\lambda \(\bar n^\mu\frac{P^+}{M}-n^\mu \frac{M}{2P^+}\)+s_T^\mu.
\end{eqnarray}
All TMD distributions are dimensionsless real functions that depend on $b^2$ (the argument $b$ is used for shortness). So, there are eight TMD distributions of twist-two.

The parameterization of the TMDFF follows the same pattern. The main difference is that operator for TMDFF is oriented along $n$ direction, and thus one should replace $n\leftrightarrow \bar n$ in (\ref{def:TMDPDF11:g+} - \ref{def:TMDPDF11:s+}). 
Also, traditionally the TMD distributions are defined in the momentum space, where it is imposed the signs of TMDPDF and TMDFF to coincide. It leads to the extra $(-1)$ factor for all distributions with odd number of $b$'s in the position space.
There are only two TMDFF for spinless hadron
\begin{eqnarray}
\label{def:TMDsFF:1:g+}
\Delta_{11}^{[\gamma^-]}(z,b)&=&D_1(z,b),
\\\label{def:TMDsFF:1:g+5}
\Delta_{11}^{[\gamma^-\gamma^5]}(z,b)&=&0,
\\\label{def:TMDsFF:1:s+}
\Delta_{11}^{[i\sigma^{\alpha-}\gamma^5]}(z,b)&=& i\epsilon_T^{\alpha\mu}b_\mu m_h H_1^\perp(z,b),
\end{eqnarray}
where $m_h$ is the mass of the produced hadron.  
We follow the standard practice \cite{Boer:1997nt, Bacchetta:2006tn} and name TMDFF analods of TMDPDFs by capital letters. The only exception is unpolarized TMDFF that is denoted by $D$. 

The parametrization for TMD distributions of twist-three is derived in ref.\cite{Rodini:2022wki}. It reads
\begin{eqnarray}\label{def:TMDsPDF:2:g+}
\Phi_{\bullet}^{\mu[\gamma^+]}(x_{1,2,3},b)&=&
\epsilon_T^{\mu\nu}s_{T\nu} M f_{\bullet T}(x_{1,2,3},b)
+ ib^\mu M^2 f^\perp_\bullet(x_{1,2,3},b)
\\\nn &&
+i\lambda \epsilon_T^{\mu\nu}b_\nu M^2 f^\perp_{\bullet L}(x_{1,2,3},b)
+b^2M^3\epsilon_T^{\mu\nu}\ta \frac{g_{T,\nu\rho}}{2}-\frac{b_\nu b_\rho}{b^2}\tc s^{\rho}_Tf_{\bullet T}^\perp(x_{1,2,3},b),
\\\label{def:TMDsPDF:2:g+5}
\Phi_{\bullet}^{\mu[\gamma^+\gamma^5]}(x_{1,2,3},b)&=&
s_T^\mu M g_{\bullet T}(x_{1,2,3},b)
-i\epsilon^{\mu\nu}_Tb_\nu M^2 g^\perp_\bullet(x_{1,2,3},b)
\\\nn &&
+i\lambda b^\mu M^2 g_{\bullet L}^\perp(x_{1,2,3},b)
+b^2M^3\ta \frac{g_T^{\mu\nu}}{2}-\frac{b^\mu b^\nu}{b^2}\tc s_{T\nu}g_{\bullet T}^\perp(x_{1,2,3},b),
\\\nn
\Phi_{\bullet}^{\mu[i\sigma^{\alpha+}\gamma^5]}(x_{1,2,3},b)&=&
\lambda g_{T}^{\mu\alpha} M h_{\bullet L}(x_{1,2,3},b) 
+\epsilon^{\mu\alpha}_T M h_\bullet(x_{1,2,3},b)
+ig_T^{\mu\alpha}(b\cdot s_T)M^2 h_{\bullet T}^{D\perp}(x_{1,2,3},b)
\\\nn &&
+i(b^\mu s^\alpha_T-s_T^\mu b^\alpha)M^2h_{\bullet T}^{A\perp}(x_{1,2,3},b)
+(b^\mu \epsilon^{\alpha\beta}_Tb_\beta+\epsilon_T^{\mu\beta}b_\beta b^\alpha)M^3h_{\bullet}^\perp(x_{1,2,3},b)
\\\label{def:TMDsPDF:2:s+} &&
+\lambda M^3 b^2\ta \frac{g^{\mu\alpha}_T}{2}-\frac{b^\mu b^\alpha}{b^2}\tc h_{\bullet L}^\perp(x_{1,2,3},b)
\\\nn &&
+i(b\cdot s_T) M^2\ta \frac{g^{\mu\alpha}_T}{2}-\frac{b^\mu b^\alpha}{b^2}\tc h_{\bullet T}^{T\perp}(x_{1,2,3},b)
\\\nn &&
+iM^2\ta \frac{b^\mu s^\alpha_T+s_T^\mu b^\alpha}{2}-\frac{b^\mu b^\alpha}{b^2}(b\cdot s_T)\tc h_{\bullet T}^{S\perp}(x_{1,2,3},b)
,
\end{eqnarray}
where $\bullet$ is $\oplus$ or $\ominus$, and $x_{1,2,3}$ is the shorthand notation for $(x_1,x_2,x_3)$. The notation for the TMD distributions follows the traditional pattern used in the parameterization of leading TMD distributions (\ref{def:TMDPDF11:g+}-\ref{def:TMDsFF:1:s+}). Namely, the proportionality to $b$ is marked by the superscript $\perp$, and the polarization by subscript $L$ (for longitudinal) or $T$ (for transverse). In the tensor case, there are four structures $\sim b^\mu s^\alpha_T$, which are denoted as $h_T^{A\perp}$, $h_T^{D\perp}$, $h_T^{S\perp}$, $h_T^{T\perp}$ for antisymmetric, diagonal, symmetric, and traceless components. 

For the TMDFFs of twist-three we same convention for naming as for twist-two case. We obtain
\begin{eqnarray}
\label{def:TMDsFF:2:g+}\Delta_{\bullet}^{\mu[\gamma^-]}(z_{1,2,3},b)&=&
-ib^\mu m_h^2 D^\perp_\bullet(z_{1,2,3},b), 
\\\label{def:TMDsFF:2:g+5}
\Delta_{\bullet}^{\mu[\gamma^-\gamma^5]}(z_{1,2,3},b)&=&
-i\epsilon^{\mu\nu}_T b_\nu m_h^2 G^\perp_\bullet(z_{1,2,3},b),
\\\label{def:TMDsFF:2:s+5}
\Delta_{\bullet}^{\mu[i\sigma^{\alpha-}\gamma^5]}(z_{1,2,3},b)&=&
-\epsilon^{\mu\alpha}_T m_h H_\bullet(z_{1,2,3},b)-(b^\mu \epsilon^{\alpha\beta}_Tb_\beta+\epsilon_T^{\mu\beta}b_\beta b^\alpha)m_h^3H_{\bullet}^\perp(z_{1,2,3},b),
\end{eqnarray}
where $z_{1,2,3}$ is the shorthand notation for $(z_1,z_2,z_3)$. 

The barred distribution $\overline{\Phi}$ and $\overline{\Delta}$ are defined by similar operators but with quark and anti-quark fields exchanged, compare (\ref{def:PHI11}-\ref{def:PHI12}) to (\ref{def:PHI11-bar}-\ref{def:PHI12-bar}). The parametrization for barred correlators is identical to usual ones, but with extra bar on the distributions letters. The unbarred $\Phi$, $\Delta$ and barred $\overline{\Phi}$, $\overline{\Delta}$ distributions are related to quark and anti-quark distributions correspondingly. So, all unbarred distributions are quark-distributions
\begin{eqnarray}
F=F_q,
\end{eqnarray}
where $F$ is TMDPDF or TMDFF of twist-two or three (in the case of TMDPDF of twist-three we also select $x_3>0$). To establish the relation between barred and anti-quark distributions one needs to inspect the C-conjugation property, see f.i.\cite{Boer:1997nt, Boer:2003cm}. It gives the relations
\begin{eqnarray}
\overline{F}=\eta[F] F_{\bar q},
\end{eqnarray}
where $\eta[F]=\pm1$ depending on the type of distribution. We found that $\eta[F]=-1$ for
\begin{eqnarray}\nn
F&\in&\{g_1, g_{1T}^\perp, 
f_{\bullet T},f_{\bullet}^\perp , f_{\bullet L}^\perp, f_{\bullet T}^\perp,
h_{\bullet L},h_{\bullet} , h_{\bullet T}^{D\perp}, h_{\bullet T}^{A\perp},
h_{\bullet}^{\perp}, h_{\bullet L}^{\perp}, h_{\bullet T}^{T\perp},
D_{\bullet}^\perp,H_{\bullet},H_{\bullet}^{\perp}
\},
\end{eqnarray}
and $\eta[F]=+1$ for all other cases. In other terms, for twist-two distributions $\eta[F^{[\Gamma]}]=-1$ for $\Gamma=\gamma^\pm\gamma^5$, while for twist-three distributions $\eta[F^{[\Gamma]}]=+1$ for $\Gamma=\gamma^\pm\gamma^5$, and opposite signs in other cases.

Despite having 32 TMDPDF of twist-three (and 8 TMDFFs) only half of them enter the factorization theorem for SIDIS. It is related to the internal joined spin of the quark-gluon pair. As for the twist-two TMD distributions, also twist-three distributions divided between T-even and T-odd distributions. The latter changes the sing upon the reorientation of the Wilson line from the SIDIS to Drell-Yan case \cite{Collins:2002kn}
\begin{eqnarray}\label{T-odd}
F(x_{1,2,3},b)\Big|_{\text{SIDIS}}=-F(x_{1,2,3},b)\Big|_{\text{DY}}.
\end{eqnarray}
The T-odd distributions that appear at NLP SIDIS are
\begin{eqnarray}
\text{T-odd}:\quad \{
f_{1T}^\perp,~ h_1^\perp,~f_{\oplus}^\perp,~g_{\ominus}^\perp, ~f_{\ominus L}^\perp, ~g_{\oplus L}^\perp,~f_{\ominus T}^\perp, ~g_{\oplus T}^\perp,~f_{\ominus T}, ~g_{\oplus T},~h_{\ominus},~h_{\ominus L},~h_{\oplus T}^{A\perp},~h_{\oplus T}^{D\perp}\},
\end{eqnarray}
and same for TMDFF analogs. As usually, we use the SIDIS definition as primary.

\subsection{Evolution}
\label{sec:evol}

TMD distributions depend on the scales $\mu$ and $\zeta$, which are hard and rapidity factorization scales. Consequently, each distribution satisfies two evolution equations \cite{Aybat:2011zv, Chiu:2012ir, Scimemi:2018xaf}. Herewith, rapidity evolution is universal for all TMD distributions at NLP. The evolution with $\mu$ is distinct for TMD distributions of twist-two, TMDPDFs of twist-three and TMDFFs of twist-three.

The evolution of TMD distributions with respect to the scale $\zeta$ is the same for all distributions of twist-two and twist-three. It reads \cite{Vladimirov:2021hdn, Ebert:2021jhy}
\begin{equation}\label{evol:rapidity}
\zeta \frac{\partial}{\partial \zeta} F(b,\zeta,\mu) = -\mathcal{D}(b,\mu)F(b,\zeta,\mu)
\end{equation}
where $\mathcal{D}(b,\mu)$ is the Collins-Soper kernel \cite{Collins:1984kg}, and $F$ is any TMD distribution of twist-two or twist-three. The Collins-Soper kernel is a universal nonperturbative function, which describes the soft-interaction in-between partons. At small-b, it can be computed perturbatively \cite{Vladimirov:2016dll} and known up to N$^4$LO \cite{Moult:2022xzt}. At large-b, it is dominated by QCD vacuum effects \cite{Vladimirov:2020umg}. The latest determinations of Collins-Soper kernel can found in refs.~\cite{Moos:2023yfa, Bacchetta:2022awv} (from the data), \cite{Shu:2023cot, LPC:2022ibr} (from the lattice simulations), \cite{BermudezMartinez:2022ctj} (from parton shower generator). 

The evolution equation with respect to parameter $\mu$ is different for twist-two and twist-three cases. All TMDPDFs and TMDFFs of twist-two evolve with the equation
\begin{eqnarray}
\mu^2\frac{d F(x,b;\mu,\zeta)}{d\mu^2}=\(\frac{\Gamma_{\text{cusp}}}{2}\ln\(\frac{\mu^2}{\zeta}\)-\frac{\gamma_V}{2}\)F(x,b;\mu,\zeta),
\end{eqnarray}
where $\Gamma_{\text{cusp}}$ is the anomalous dimension of light-like cusp of Wilson lines, and $\gamma_V$ is the anomalous dimension associated with the vector form-factor of the quark. At LO, these anomalous dimensions are $\Gamma_{\text{cusp}}=4C_Fa_s$ and $\gamma_V=-6C_Fa_s$.

The evolution of TMDPDFs of twist-three is rather involved. The full description and derivation of the following equation is given in ref.\cite{Rodini:2022wki}. 
The main complication arises from the sign-indefiniteness of variables $x_i$. Due to it, a part of anomalous dimension (for distributions $\Phi_{12}$ and $\Phi_{21}$) is complex, and proportional to the direction of the Wilson line. 
In the terms of distributions $\Phi_{\oplus}$ and $\Phi_{\ominus}$, the complex part transforms into a mixing term. The evolution equation reads
\begin{eqnarray}\label{def:evol-PDF}
\mu^2 \frac{d}{d\mu^2}\(\begin{array}{c}F_1\\ F_2\end{array}\) &=&
\(\frac{\Gamma_{\text{cusp}}}{2}\ln\(\frac{\mu^2}{\zeta}\)
+\Upsilon_{x_1x_2x_3}\)
\(\begin{array}{c}F_1\\ F_2\end{array}\)
\\\nn &&\qquad\qquad
+
\(\begin{array}{cc}
2\mathbb{P}_A &  2\pi \Theta_{x_1x_2x_3}\\
-2\pi\Theta_{x_1x_2x_3}     & 2\mathbb{P}_A
\end{array}\)\(\begin{array}{c}F_1\\ F_2\end{array}\),
\end{eqnarray}
where all distributions are functions of $(x_1,x_2,x_3,b;\mu,\zeta)$. The functions $\Upsilon$ and $\Theta$ are multiplicative, while the kernel $\mathbb{P}_A$ is the integral kernel that acts on the TMDPDF. The function $\Theta$ is discontinuous at $x_i=0$, and the function $\Upsilon$ has logarithmic singularities in these points. The LO expressions for these elements are collected in appendix \ref{app:evolution}. The pairs $(F_1,F_2)$ could be any of the following
\begin{eqnarray}\label{F1F2}
\(\begin{array}{c}f_{\oplus L}+g_{\ominus L}\\ f_{\ominus L}-g_{\oplus L}\end{array}\),
\quad
\(\begin{array}{c}f_{\oplus}^\perp+g_{\ominus}^\perp\\ f_{\ominus}^\perp-g_{\oplus}^\perp\end{array}\),
\quad
\(\begin{array}{c}f_{\oplus L}^\perp+g_{\ominus L}^\perp\\ f_{\ominus L}^\perp-g_{\oplus L}^\perp\end{array}\),
\quad
\(\begin{array}{c}f_{\oplus T}^\perp+g_{\ominus T}^\perp\\ f_{\ominus T}^\perp-g_{\oplus T}^\perp\end{array}\),
\\\nn
\(\begin{array}{c}h_\oplus\\ h_\ominus\end{array}\),
\quad
\(\begin{array}{c}h_{\oplus L}\\ h_{\ominus L}\end{array}\),
\quad
\(\begin{array}{c}h_{\oplus T}^{D\perp}\\ h_{\ominus T}^{D\perp}\end{array}\),
\quad
\(\begin{array}{c}h_{\oplus T}^{A\perp}\\ h_{\ominus T}^{A\perp}\end{array}\).
\end{eqnarray}
Other functions (and orthogonal combinations of the ones above) evolve with the evolution kernel $\mathbb{P}_B$ \cite{Rodini:2022wki}, and do not appear in SIDIS.

Let us mention a peculiar feature of evolution equation (\ref{def:evol-PDF}). In the pair $(F_1,F_2)$ (\ref{F1F2}) exactly one component is T-odd, and one is T-even. This feature seems to be inconsistent the T-invariance of QCD, because of mixing terms in (\ref{def:evol-PDF}). Nonetheless, the T-invariance is preserved, because for the TMDPDFs with Wilson lines pointing in the other direction (i.e. for DY-defined TMDPDFs) the function $\Theta$ changes the sign. Therefore, ``modified'' universality of naively T-odd TMD distribution functions (\ref{T-odd}) is preserved.

The evolution equation for the TMDFFs of twist-three has simpler structure. It is due to the impossibility to mix the regions with different signs of $z_i$ (\ref{support:Delta}). For that reason the anomalous dimension is strictly real, and components $\Delta_\oplus$ and $\Delta_\ominus$ evolve independently, and with the same equation. The evolution equation reads
\begin{eqnarray}\label{def:TMDFF-evol}
\mu^2 \frac{d}{d\mu^2}\Delta=
\Big[
\frac{\Gamma_{\text{cusp}}}{2}\ln\(\frac{\mu^2}{\zeta}\)+\Upsilon_{z_1z_2z_3}^{\text{FF}}+\mathbb{P}_A^{\text{FF}}\Big]\Delta,
\end{eqnarray}
where we omit the argument $(z_1,z_2,z_3,b;\mu,\zeta)$ for TMDPDF, $\Gamma_{\text{cusp}}$ is the cusp anomalous dimension, $\Upsilon^{\text{FF}}$ is the multiplicative factor, and $\mathbb{P}_A^{\text{FF}}$ is the integral kernel. The LO expressions for these terms are presented in the appendix \ref{app:evolution}. The equation (\ref{def:TMDFF-evol}) is valid only for TMDFF belonging to the following set
\begin{eqnarray}
\Delta\in \{
D_{\oplus}^\perp-G_{\ominus}^\perp, D_{\ominus}^\perp+G_{\oplus}^\perp,
H_\oplus, H_\ominus\}.
\end{eqnarray}
For the rest combinations (\ref{app:DELTAB}) the kernel $\mathbb{P}_A^{\text{FF}}$ must be replaced by $\mathbb{P}_B^{\text{FF}}$ (\ref{app:PB-FF}). The evolution kernel $\mathbb{P}_A$ were also derived in ref.\cite{Beneke:2017ztn}, and agrees with our expressions.

The combinations $f_\oplus^\perp+g^\perp_\ominus$ and $f^\perp_{\ominus}-g^\perp_\oplus$ (and similar) can be considered as a single nonperturbative function. These combination not only evolve as a single object, but also appear in the factorization theorem. Also exactly these combinations appear in the relations between dynamical  and geometrical twist-three distributions. The complete set of relations is given in ref.\cite{Rodini:2022wki} eqns.(6.11)-(6.26).

\section{Differential cross-section for SIDIS at NLP}
\label{sec:SIDIS}

The SIDIS is the reaction
\begin{eqnarray}
\ell(l)+N(P)\to \ell(l')+h(p_h)+X,
\end{eqnarray}
where $\ell$ is the lepton, $N$ is the nucleon target, and $h$ is the produced hadron. The four-momenta of particles are indicated in brackets. The SIDIS process is considered in many works. The canonical paper that systematized the structure functions of polarized SIDIS is ref.\cite{Bacchetta:2006tn}. We use this paper as the main reference. 

In the following sections we present the computation and the result of the cross-section for the polarized SIDIS at NLP of TMD factorization theorem, using the hadronic tensor presented in sec.\ref{sec:NLP-gen}. The main technical difficulty in this computation is the transformation between laboratory frame (where the structure functions are defined), and the factorization frame (where the factorization theorem is derived).

\begin{figure}[t]
\centering
\includegraphics[width=0.65\textwidth]{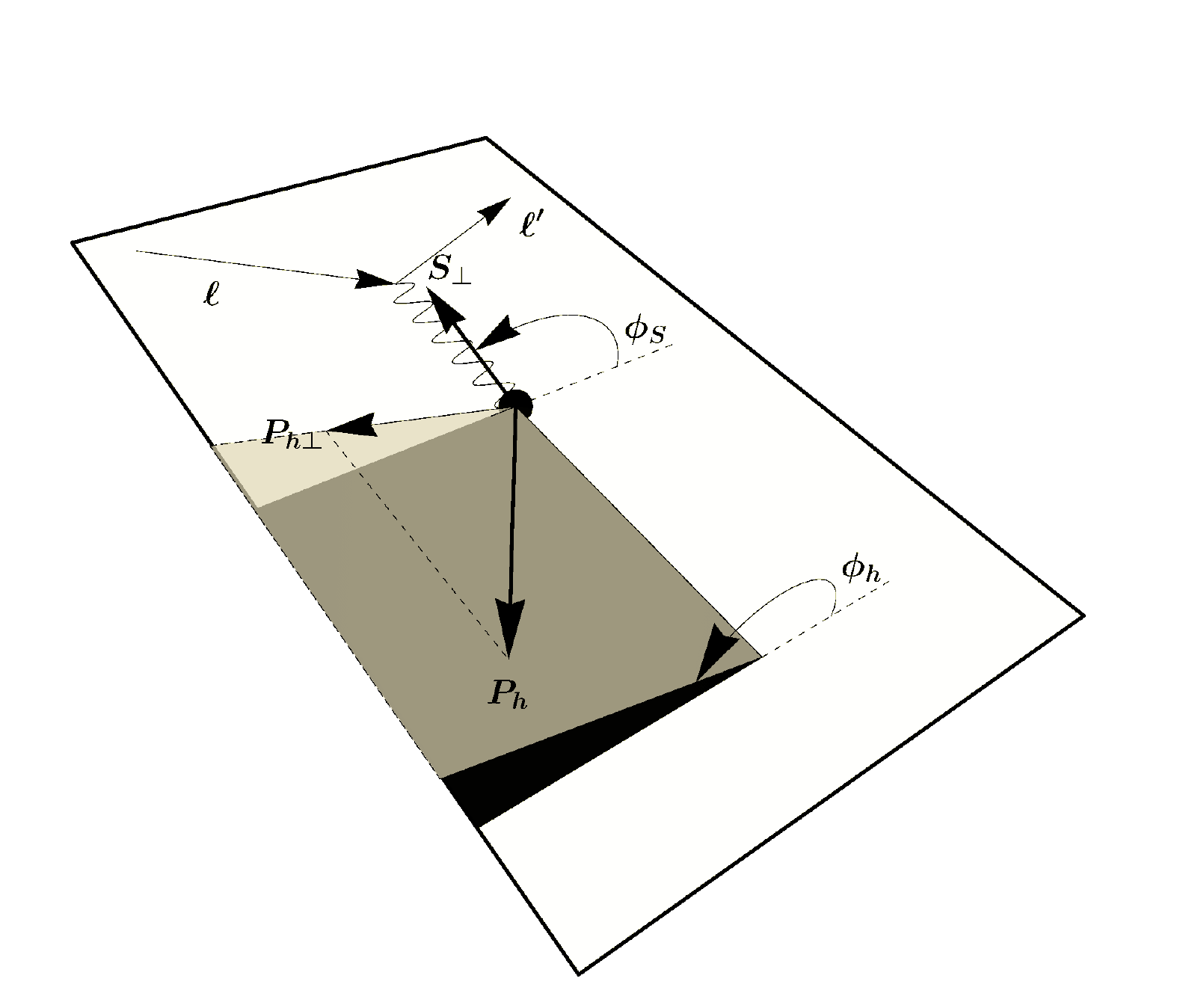}
\caption{Definition of azimuthal angles for SIDIS in the laboratory frame. The subscript $\perp$ labels the transverse components of corresponding vectors.}
\label{fig:trento}
\end{figure}

\subsection{Kinematics of SIDIS}
\label{sec:SIDIS-kinematics}

The main kinematic variables of the SIDIS are
\begin{eqnarray}
x=\frac{Q^2}{2(qP)},\qquad y=\frac{(qP)}{(lP)},\qquad z=\frac{(p_hP)}{(qP)},\qquad \gamma=\frac{2Mx}{Q},\qquad \gamma_h=\frac{m_h}{zQ},
\end{eqnarray}
where $q=l-l'$, $Q^2=-q^2>0$, $M^2=P^2$ is the mass of target hadron, $m_h^2=p_h^2$ is the mass of produces hadron and the leptons are considered massless, $l^2=l'^2=0$.  We emphasize that the factorized expression derived in this paper is valid up to NLP. I.e. it is valid up to corrections $\sim M^2/Q^2$, $\sim m_h^2/Q^2$, $\sim p_{h\perp}^2/Q^2$, etc. Therefore, in the final expression we systematically drop suppressed terms. However, in order to obtain corresponding expressions, we start from the complete kinematic picture, which includes $\gamma$, $\gamma_h$, and $p_\perp$ terms, and drop the power suppressed terms after the tensor convolutions. It is necessary because some combinations have naively singular or naively vanishing behavior at vanishing masses. Following such a non-minimal procedure we safely recover all NLP corrections.

The azimuthal plane is defined as orthogonal to vectors $q^\mu$ and $P^\mu$. The components of vectors belonging to this plane are denoted by subscript $\perp$. The corresponding tensors are
\begin{eqnarray}
g_\perp^{\mu\nu}&=&g^{\mu\nu}-\frac{1}{Q^2(1+\gamma^2)}\qa 4x^2 P^\mu P^\nu - \gamma^2 q^\mu q^\nu +2x (P^\mu q^\nu + q^\mu P^\nu)\qc,
\\
\epsilon_\perp^{\mu\nu}&=&\epsilon^{\mu\nu\rho\sigma}\frac{2x P_\rho q_\sigma}{Q^2\sqrt{1+\gamma^2}}.
\end{eqnarray}
We use the definition $\epsilon^{0123}=+1$, which leads to $\epsilon_\perp^{12}=\epsilon_\perp^{21}=1$. The spin-vector of the target is decomposed as
\begin{eqnarray}
S^\mu=S_\|\(\frac{P^\mu}{Q}\frac{2x}{\gamma\sqrt{1+\gamma^2}}-\frac{q^\mu}{Q}\frac{\gamma}{\sqrt{1+\gamma^2}}\)+S_\perp^\mu,
\end{eqnarray}
where $S^\mu_\perp=g_\perp^{\mu\nu}S_\nu$.

The azimuthal angles between vectors are defined according to the Trento convention \cite{Bacchetta:2004jz}, and shown in fig.\ref{fig:trento}. They can be computed using the tensors $g_\perp$ and $\epsilon_\perp$. The azimuthal angle for the produced hadron is defined as
\begin{eqnarray}
\cos\phi_h=\frac{-(lp_h)_\perp}{\sqrt{l_\perp^2 p_{h\perp}^2}},
\qquad
\sin\phi_h=\frac{-l_\mu p_{h\nu}\epsilon^{\mu\nu}_\perp}{\sqrt{l_\perp^2 p_{h\perp}^2}},
\end{eqnarray}
where $(ab)_\perp=a_\mu b_\nu g^{\mu\nu}_\perp$. The azimuthal angle for spin-vector is similarly defined as
\begin{eqnarray}
\cos\phi_S=\frac{-(lS)_\perp}{\sqrt{l_\perp^2 S_{\perp}^2}},
\qquad
\sin\phi_S=\frac{-l_\mu S_{\nu}\epsilon^{\mu\nu}_\perp}{\sqrt{l_\perp^2 S_{\perp}^2}}.
\end{eqnarray}

The cross-section for the SIDIS in the one-photon approximation reads
\begin{eqnarray}
d\sigma=\frac{2}{s-M^2}\frac{\alpha_{\text{em}}^2}{Q^4}\frac{d^3l'}{2E'}\frac{d^3p_h}{2E_h} L_{\mu\nu}W^{\mu\nu},
\end{eqnarray}
where $E'$ and $E_h$ are energies of final-state lepton and produced hadrons, respectively. The letonic tensor $L^{\mu\nu}$ is
\begin{eqnarray}
L_{\mu\nu}=2(l_\mu l'_{\nu}+l'_\mu l_{\nu}-(ll')g_{\mu\nu})+2i\lambda_e \epsilon_{\mu\nu\rho\sigma}l^\rho l'^{\sigma},
\end{eqnarray}
where $\lambda_e$ is the helicity of the lepton beam. The hadronic tensor (in the TMD factorization) is given in eqn.(\ref{W:W-all}). In the conventional variables the differential cross-section is
\begin{eqnarray}\label{xSec=LW}
\frac{d\sigma}{dx\,dy\,dz\,d\phi_h\,d\phi_S\,dp_\perp^2}=\frac{\alpha^2_{\text{em}}}{Q^4}\frac{y}{8z}L_{\mu\nu}W^{\mu\nu},
\end{eqnarray}
where we have dropped the corrections of order $\mathcal{O}(\gamma^2\gamma_h^2)$. The decomposition of the cross-section over the structure functions\footnote{
We omit the factor $(1+\gamma^2/2x)$ that is present in the original expression in ref. \cite{Bacchetta:2006tn}. This factor does not follow from any invariant decomposition of tensors, and was included in ref.\cite{Bacchetta:2006tn} in order to match the normalization of the integrated cross-section, see also \cite{Boglione:2019nwk}. In any case, this term is  N$^2$LP and thus irrelevant for the present work.
} is \cite{Bacchetta:2006tn}
\begin{eqnarray}\label{xSec=F}
\frac{d\sigma}{dxdy dz d\phi_h d\phi_S dp_\perp^2}
&=&
\frac{\alpha_{em}^2}{xy Q^2}\frac{y^2}{2(1-\varepsilon)}\Bigg\{
F_{UU,T}+\varepsilon F_{UU,L}+\sqrt{2\varepsilon(1+\varepsilon)}\cos \phi_h F_{UU}^{\cos \phi_h}
\\\nn &&
+\varepsilon \cos(2\phi_h) F_{UU}^{\cos 2\phi_h}
+\lambda_e \sqrt{2\varepsilon(1-\varepsilon)}\sin\phi_h F_{LU}^{\sin\phi_h}
\\\nn &&
+S_{\|}\Big[\sqrt{2\varepsilon(1+\varepsilon)}\sin \phi_h F_{UL}^{\sin \phi_h}+\varepsilon \sin(2\phi_h)F_{UL}^{\sin 2\phi_h}
\Big]
\\\nn &&
+S_{\|}\lambda_e\Big[\sqrt{1-\varepsilon^2}F_{LL}+\sqrt{2\varepsilon(1-\varepsilon)}\cos\phi_h F_{LL}^{\cos\phi_h}\Big]
\\\nn &&
|S_\perp|\Big[\sin(\phi_h-\phi_S)\(F_{UT,T}^{\sin(\phi_h-\phi_S)}+\varepsilon F_{UT,L}^{\sin(\phi_h-\phi_S)}\)
\\\nn &&
+\varepsilon \sin(\phi_h+\phi_S)F_{UT}^{\sin(\phi_h+\phi_S)}
+\varepsilon \sin(3\phi_h-\phi_S)F_{UT}^{\sin(3\phi_h-\phi_S)}
\\\nn &&
+\sqrt{2\varepsilon(1+\varepsilon)}\sin\phi_S F_{UT}^{\sin \phi_S}
+\sqrt{2\varepsilon(1+\varepsilon)}\sin(2\phi_h-\phi_S)F_{UT}^{\sin(2\phi_h-\phi_S)}\Big]
\\\nn &&
+ |S_\perp|\lambda_\epsilon \Big[
\sqrt{1-\varepsilon^2}\cos(\phi_h-\phi_S)F_{LT}^{\cos(\phi_h-\phi_S)}
+\sqrt{2\varepsilon(1-\varepsilon)}\cos\phi_SF_{LT}^{\cos\phi_S}
\\\nn &&
+\sqrt{2\varepsilon(1-\varepsilon)}\cos(2\phi_h-\phi_S)F_{LT}^{\cos(2\phi_h-\phi_S)}\Big]\Bigg\},
\end{eqnarray}
where
\begin{eqnarray}
\varepsilon=\frac{1-y-\frac{\gamma^2y^2}{4}}{1-y+\frac{y^2}{2}+\frac{\gamma^2y^2}{4}}.
\end{eqnarray}
Note, that in this decomposition the terms $\sim \gamma_h$ are omitted. We have checked explicitly that at NLP such terms do not appear. The apparent asymmetry in the treatment of $\gamma$ and $\gamma_h$ comes from working in the laboratory frame.

The TMD factorization is derived in the frame where the target and produced hadrons are back-to-back \cite{Collins:2011zzd}. This is a critical assumption because it allows the mode separation and counting rules for the field components, see f.i. \cite{Vladimirov:2021hdn, Ebert:2021jhy}. In the present approach to the TMD factorization, the factorization frame could not be defined for massive hadrons. This would require accounting the target-mass corrections, and for the moment they cannot be incorporated into the factorization formalism. Moreover, the factorization frame is defined ambiguously, since the vectors $n$ and $\bar n$ could be modified by power corrections, the so-called reparametrization invariance \cite{Chay:2002vy, Manohar:2002fd}. Both these effects are of N$^2$LP nature, and thus they do not affect our computation. Simultaneously, it grants as a freedom in the definition of the factorization frame. We define
\begin{eqnarray}\label{def:n-nbar}
n^\mu &=& \frac{2x P^+}{zQ^2\sqrt{1-\gamma^2\gamma_h^2}}\(p_h^\mu -P^\mu \frac{2xz(1-\sqrt{1-\gamma^2 \gamma_h^2})}{\gamma^2}\) = \frac{2xP^+}{zQ^2}p_h^\mu +\mathcal{O}(\text{N$^2$LP}),
\\
\bar n^\mu &=& \frac{1}{2P^+\sqrt{1-\gamma^2\gamma_h^2}}\(P^\mu (1+\sqrt{1-\gamma^2 \gamma_h^2})
-p_h^\mu\frac{\gamma^2}{2xz}\)= \frac{P^\mu}{P^+} +\mathcal{O}(\text{N$^2$LP}),
\end{eqnarray}
where on the right-hand-side we present the leading term in the power decomposition. These definitions satisfy $n^2=\bar n^2=0$, and $(n\bar n)=1$, and the vectors aligned along the large component of the hadrons' momenta. The component $P^+$ in these expressions remain undefined. This is not an issue since it necessarily cancels in all final expressions due to the boost-invariance.

The hadron tensor in the TMD factorization is naturally described in the terms of transverse components relative to the plane $(n, \bar n)$. By definition of (\ref{def:n-nbar}) this plain coincides with the plane $(P,p_h)$ and thus the corresponding tensors  (\ref{def:gT}) are orthogonal to $P^\mu$ and $p_h^\mu$,
\begin{eqnarray}
g_T^{\mu\nu}&=&g^{\mu\nu}-\frac{1}{Q^2(1-\gamma^2\gamma_h^2)}\[
\frac{2x}{z}(P^\mu p_h^\nu+p_h^\mu P^\nu)
-\frac{\gamma^2}{z^2}p_h^\mu p_h^\nu
-4x^2\gamma_h^2 P^\mu P^\nu\],
\\
\epsilon_T^{\mu\nu}&=&\frac{2x}{zQ^2\sqrt{1-\gamma^2\gamma_h^2}}\epsilon^{\mu\nu \sigma\rho}P_\alpha p_{h\rho}.
\end{eqnarray}
Comparing various invariants in terms of T-components and $\perp$-components we find the translation dictionary between frames. 

As it was pointed before, the relations between various components are ambiguous at N$^2$LP, therefore, there is no reason to present them here. It is only important to mention the key combinations of the factorization expression
\begin{eqnarray}
\frac{q^+}{P^+}= -x+\mathcal{O}(\text{N$^2$LP}),
\qquad
\frac{p_h^-}{q^-}= z+\mathcal{O}(\text{N$^2$LP}),
\qquad
q_T^2= \frac{p_\perp^2}{z^2}+\mathcal{O}(\text{N$^2$LP}).
\end{eqnarray}
Due to it the variables $\tilde x$ and $\tilde z$ in (\ref{W:W-all}) are equal to $x$ and $z$ and the present accuracy.

\subsection{Expressions for the structure functions are NLP}
\label{sec:final-expression}

The computation of structure functions is lengthy but straightforward. We contract hadronic and leptonic tensors (\ref{xSec=LW}) and obtain the expression in the terms of factorization variables, i.e. defined by $g_T^{\mu\nu}$, $n^\mu$, $\bar n^\mu$. Next, we rewrite these scalar products in terms of standard variables, and expand the obtained expression at $Q\to\infty$ up to the NLP term. Since the TMD distributions depend on $b^2$ we integrate over the angular part of $b^\mu$, and express all Fourier transforms via Hankel trnaforms. Finally, comparing with the parametrization (\ref{xSec=F}) we obtain expressions for structure functions. The algebraic operations are rather bulky and were performed with the help of \textit{FeynCalc} package \cite{Shtabovenko:2020gxv}. 

As result of this computation we obtain four type of terms. 
\begin{itemize}
\item[\textit{(i)}] The leading contribution of $W_{\text{LP}}$ (\ref{def:W-LP}). It is $\sim Q^0$, and produces the well-known LP TMD factorization.
\item[\textit{(ii)}] The NLP contribution of $W_{\text{LP}}$ (\ref{def:W-LP}). It is $\sim |p_\perp|/(z Q)$. These terms appears due to the expansion of scalar products in the factorization kinematics, and contain only TMD distributions of twist-two.
\item[\textit{(iii)}] The leading contribution of $W_{\text{kNLP}}$ (\ref{def:Wk}). It is $\sim M/Q$, and contains TMD distributions of twist-two, their derivatives, and the derivatives of Collins-Soper kernel.
\item[\textit{(iv)}] The leading contribution of $W_{\text{gNLP}}$ (\ref{def:Wk}). It is $\sim M/Q$, and contains TMD distributions of twist-two and twist-three.
\end{itemize}
The terms \textit{(ii)} and \textit{(iii)} have the same hard coefficient function, and could be combined together. Moreover, their structure mutually simplifies due to integration by parts. Since these terms present a theoretical interest we collected expressions for \textit{(ii)} and \textit{(iii)} in appendix \ref{app:KPC}, before simplification.

All terms have the form of Hankel transform of a product of two distributions accompanied by the coefficient function. To have a compact expression for the result, we introduce the short-hand notation for such combination. So, the terms of types $(i)$, $(ii)$ and $(iii)$ are expressed via
\begin{eqnarray}\label{def:JN}
\mathcal{J}_n[fD]&=&x|C_1(\mu^2,Q^2)|^2 \sum_i e_i^2 \int \frac{b db}{2\pi} (bM)^n J_n\(\frac{b|p_\perp|}{z}\) f_{i}(x,b;\mu,\zeta)D_i(z,b;\mu,\bar \zeta),
\end{eqnarray}
where $J_n$ is the Bessel function of the order $n$, $f$ is a TMDPDF, $D$ is a TMDFF, $e_i$ is the electric charge of the quark with flavor $i$. The sum over flavors $i$ runs through quark and anti-quark flavors. The coefficient function is given in eqn.~(\ref{def:C1^2}). The terms of type $(iv)$ have three kinds of structures
\begin{eqnarray}
\mathcal{J}^{[2]}_n[fD_\bullet]&=&x\sum_i e_i^2  \int \frac{b db}{2\pi} (bM)^n J_n\(\frac{b|p_\perp|}{z}\)\\\nn &&\times
\int_{-\infty}^{-z}dw_2 \frac{z|w_2|}{|w_2|-z} \mathbb{C}_2(z,w_2)  f_i(x,b;\mu,\zeta)D_{i\bullet}\(\frac{-w_2z}{w_2+z},w_2,z,b;\mu,\bar \zeta\),
\\\nn
\mathcal{J}^{[R]}_n[f_\bullet D]&=&x\sum_i e_i^2  \int \frac{b db}{2\pi} (bM)^n J_n\(\frac{b|p_\perp|}{z}\)\\\nn &&\times
\int_{-1}^{1-x}du_2 \mathbb{C}_R(x,u_2)  f_{i\bullet}(-x-u_2,u_2,x,b;\mu,\zeta)D_i(z,b;\mu,\bar \zeta),
\\\nn
\mathcal{J}^{[I]}_n[f_\bullet D]&=&x\sum_i e_i^2  \int \frac{b db}{2\pi} (bM)^n J_n\(\frac{b|p_\perp|}{z}\)\\\nn &&\times
\int_{-1}^{1-x}du_2 \pi\mathbb{C}_I(x,u_2)  f_{i\bullet}(-x-u_2,u_2,x,b;\mu,\zeta)D_i(z,b;\mu,\bar \zeta),
\end{eqnarray}
where $f$($D$) is a TMDPDF (TMDFF) of twist-two and $f_\bullet$ ($D_\bullet$) is a TMDPDF (TMDFF) of twist-three. The coefficient functions are given in eqns.~(\ref{CR}, \ref{CI}, \ref{C2}). There is a freedom to define the convolution for twist-three distributions by changing the integration variable. For definiteness we integrate over the gluon momentum fraction. Note, that the integral over $w_2$ for $\mathcal{J}^{[2]}_n$ is convergent at $w_2\to-z$ since TMDFFs vanish once first argument approach zero (\ref{support:TMDFF=0}). Also note, that the convolutions $\mathcal{J}^{[R]}_n$ and $\mathcal{J}^{[I]}_n$ include positive and negative ranges of $u_2$, and regular at $u_2=0$.

The final expressions for the structure functions are
\begin{eqnarray}\label{result:FUUT}
F_{UU,T}&=&\mathcal{J}_0[f_1D_1]
\\
F_{UU,L}&=&0
\\
F_{UU}^{\cos \phi}&=&\frac{2M}{Q}\Big\{\mathcal{J}_1\big[
\mathring{f}_1D_1-2h_1^\perp H_1^\perp-M^2|b|^2\mathring{h}_1^\perp H_1^\perp \big]
\\\nn &&\qquad
+\mathcal{J}^{(2)}_1\big[f_1(D^\perp_{\ominus}+G^\perp_{\oplus})+2h_1^\perp H_\ominus\big]
\\\nn &&\qquad
-\mathcal{J}^{(R)}_1\big[(f^\perp_\ominus-g^\perp_\oplus)D_1+2h_\ominus H_1^\perp\big]
\\\nn &&\qquad
+\mathcal{J}^{(I)}_1\big[(f^\perp_\oplus+g^\perp_\ominus)D_1+2h_\oplus H_1^\perp\big]\Big\}
\\
F_{UU}^{\cos 2\phi}&=&\mathcal{J}_2[h_1^\perp H_1^\perp]
\\
F_{LU}^{\sin\phi}&=&
\frac{2M}{Q}\Big\{
-\mathcal{J}^{(2)}_1\big[f_1(D^\perp_{\oplus}-G^\perp_{\ominus})+2h_1^\perp H_\oplus\big]
\\\nn &&\qquad
+\mathcal{J}^{(R)}_1\big[(f^\perp_\oplus+g^\perp_\ominus)D_1+2h_\oplus H_1^\perp\big]
\\\nn &&\qquad
+\mathcal{J}^{(I)}_1\big[(f^\perp_\ominus-g^\perp_\oplus)D_1+2h_\ominus H_1^\perp\big]
\Big\}
\\
F_{UL}^{\sin \phi}&=&-\frac{2M}{Q}\Big\{
\mathcal{J}_1[2h_{1L}H_1^\perp+M^2|b|^2\mathring{h}_{1L}H_1^\perp]
\\\nn &&\qquad
+\mathcal{J}^{(2)}_1\big[
g_1(D^\perp_{\oplus}-G^\perp_{\ominus})-2h_{1L}^\perp H_\ominus\big]
\\\nn &&\qquad
+\mathcal{J}^{(R)}_1\big[(f^\perp_{\ominus L}-g^\perp_{\oplus L})D_1+2h_{\ominus L} H_1^\perp\big]
\\\nn &&\qquad
-\mathcal{J}^{(I)}_1\big[(f^\perp_{\oplus L}+g^\perp_{\ominus L})D_1+2h_{\oplus L} H_1^\perp\big]
\Big\}
\\
F_{UL}^{\sin 2\phi}&=&\mathcal{J}_2[h_{1L}H_1^\perp]
\\
F_{LL}&=&\mathcal{J}_0[g_{1}D_1]
\\\label{result:FLL:1}
F_{LL}^{\cos\phi}&=&\frac{2M}{Q}\Big\{
\mathcal{J}_1[\mathring{g}_{1}D_1]
\\\nn &&\qquad
+\mathcal{J}^{(2)}_1\big[
g_1(D^\perp_{\ominus}+G^\perp_{\oplus})+2h_{1L}^\perp H_\oplus\big]
\\\nn &&\qquad
-\mathcal{J}^{(R)}_1\big[(f^\perp_{\oplus L}+g^\perp_{\ominus L})D_1+2h_{\oplus L} H_1^\perp\big]
\\\nn &&\qquad
-\mathcal{J}^{(I)}_1\big[(f^\perp_{\ominus L}-g^\perp_{\oplus L})D_1+2h_{\ominus L} H_1^\perp\big]
\Big\}
\\
F_{UT,T}^{\sin(\phi-\phi_S)}&=&-\mathcal{J}_1[f_{1T}^\perp D_1]
\\
F_{UT,L}^{\sin(\phi-\phi_S)}&=&0
\\
F_{UT}^{\sin(\phi+\phi_S)}&=&\mathcal{J}_1[h_1H_1^\perp]
\\
F_{UT}^{\sin(3\phi-\phi_S)}&=&\frac{1}{4}\mathcal{J}_3[h_{1T}^\perp H_1^\perp]
\\\label{result:FUT:S}
F_{UT}^{\sin \phi_S}&=&\frac{2M}{Q}\Big\{
-\mathcal{J}_0\[f_{1T}^\perp D_1+\frac{M^2|b^2|}{2}\mathring{f}_{1T}^\perp D_1
+M^2|b|^2 \mathring{h}_{1} H_1^\perp\]
\\\nn &&\qquad
+\mathcal{J}^{(2)}_0\big[
\frac{|b|^2 M^2}{2} g_{1T}^\perp(D^\perp_{\oplus}-G^\perp_{\ominus})
-\frac{|b|^2M^2}{2}f_{1T}^\perp (D^\perp_{\ominus}+G^\perp_{\oplus})
+2h_{1} H_\ominus\big]
\\\nn &&\qquad
+\mathcal{J}^{(R)}_0\big[(f_{\ominus T}-g_{\oplus T})D_1
+|b|^2M^2 h_{\ominus T}^{A\perp} H_1^\perp
+|b|^2M^2 h_{\ominus T}^{D\perp} H_1^\perp\big]
\\\nn &&\qquad
-\mathcal{J}^{(I)}_0\big[
(f_{\oplus T}+g_{\ominus T})D_1
+|b|^2M^2 h_{\oplus T}^{A\perp} H_1^\perp
+|b|^2M^2 h_{\oplus T}^{D\perp} H_1^\perp
\big]\Big\}
\\
F_{UT}^{\sin(2\phi-\phi_S)}&=&-\frac{M}{Q}\Big\{
\mathcal{J}_2\[\mathring{f}_{1T}^\perp D_1+2h^\perp_{1T} H_1^\perp+\frac{M^2|b|^2}{2} \mathring{h}_{1T}^\perp H_1^\perp\]
\\\nn &&\qquad
+\mathcal{J}^{(2)}_2\big[
g_{1T}^\perp(D^\perp_{\oplus}-G^\perp_{\ominus})
+f_{1T}^\perp (D^\perp_{\ominus}+G^\perp_{\oplus})
-h_{1T}^\perp H_\ominus\big]
\\\nn &&\qquad
+\mathcal{J}^{(R)}_2\big[(f^\perp_{\ominus T}-g^\perp_{\oplus T})D_1
-2 h_{\ominus T}^{A\perp} H_1^\perp
+2 h_{\ominus T}^{D\perp} H_1^\perp\big]
\\\nn &&\qquad
-\mathcal{J}^{(I)}_2\big[
(f^\perp_{\oplus T}+g^\perp_{\ominus T})D_1
-2 h_{\oplus T}^{A\perp} H_1^\perp
+2 h_{\oplus T}^{D\perp} H_1^\perp\big]
\Big\}
\\
F_{LT}^{\cos(\phi-\phi_S)}&=&\mathcal{J}_1[g_{1T}^\perp D_1]
\\
F_{LT}^{\cos\phi_S}&=&-\frac{M}{Q}\Big\{\mathcal{J}_0[2g_{1T}^\perp D_1+M^2|b|^2 \mathring{g}_{1T}D_1]
\\\nn &&\qquad
+\mathcal{J}^{(2)}_0\big[
|b|^2 M^2 f_{1T}^\perp(D^\perp_{\oplus}-G^\perp_{\ominus})
+|b|^2M^2g_{1T}^\perp (D^\perp_{\ominus}+G^\perp_{\oplus})
-4h_{1} H_\oplus\big]
\\\nn &&\qquad
-2\mathcal{J}^{(R)}_0\big[(f_{\oplus T}+g_{\ominus T})D_1
+|b|^2M^2 h_{\oplus T}^{A\perp} H_1^\perp
+|b|^2M^2 h_{\oplus T}^{D\perp} H_1^\perp\big]
\\\nn &&\qquad
-2\mathcal{J}^{(I)}_0\big[
(f_{\ominus T}-g_{\oplus T})D_1
+|b|^2M^2 h_{\ominus T}^{A\perp} H_1^\perp
+|b|^2M^2 h_{\ominus T}^{D\perp} H_1^\perp\big]
\Big\}
\\\label{result:FLT:cos2S}
F_{LT}^{\cos(2\phi-\phi_S)}&=&\frac{M}{Q}\Big\{\mathcal{J}_2[\mathring{g}_{1T} D_1]
\\\nn &&\qquad
-\mathcal{J}^{(2)}_2\big[
f_{1T}^\perp(D^\perp_{\oplus}-G^\perp_{\ominus})
-g_{1T}^\perp (D^\perp_{\ominus}+G^\perp_{\oplus})
-h_{1T}^\perp H_\oplus\big]
\\\nn &&\qquad
-\mathcal{J}^{(R)}_2\big[
(f_{\oplus T}^\perp+g_{\ominus T}^\perp)D_1
-2 h_{\oplus T}^{A\perp} H_1^\perp
+2 h_{\oplus T}^{D\perp} H_1^\perp\big]
\\\nn &&\qquad
-\mathcal{J}^{(I)}_2\big[
(f_{\ominus T}^\perp-g_{\oplus T}^\perp)D_1
-2 h_{\ominus T}^{A\perp} H_1^\perp
+2 h_{\ominus T}^{D\perp} H_1^\perp\big]
\Big\}
,
\end{eqnarray}
where 
\begin{eqnarray}\label{def:mathring}
\mathring{f}(x,b;\mu,\zeta)=\frac{2}{M^2}\[\frac{\partial }{\partial |b|^2}+\frac{1}{2}\ln\(\frac{\zeta}{\bar \zeta}\)\(\frac{\partial \mathcal{D}(b,\mu)}{\partial |b|^2}\)\]f(x,b;\mu,\zeta),
\end{eqnarray}
is the dimensionless boost-invariant derivative of the TMDPDF. In these expressions we set $m_h=M$ in the definition of TMDFFs for shortness. To restore the proper coefficients one should multiply TMDFFs by factors $(m_h/M)^n$ where $n$ is (1,1,2,2,3) for $(H_1^\perp, H_\bullet, D_\bullet^\perp, G_\bullet^\perp, H_\bullet^\perp)$, correspondingly.

The structure functions related to the longitudinaly polarized photons $F_{UU,L}$ and $F_{UT,L}^{\sin(\phi-\phi_s)}$ are of N$^2$LP.

The expressions for structure functions is given in the position space, where the TMD phenomenology is usually performed. The passage to the momentum space is straightforward (see f.i.\cite{Boer:2011xd, Ebert:2021jhy}). The position space presentation is not unique for the kinematic power correction since the derivative term can be integrated by parts and presented in different ways. In our derivation we employed the strategy to eliminate the terms $\sim p_\perp$, that are produced by the LP hadronic tensor. In appendix \ref{app:KPC} we present these terms independently. The final expression is simpler expressed with the derivatives of TMDPDFs, and this is a consequence of the definition of the azimuthal angles $\phi$ and $\phi_S$.

\subsection{Discussion}

The LP structure functions, namely $F_{UU,T}$, $F_{UU}^{\cos\phi}$, $F_{UL}^{\sin 2\phi}$, $F_{LL}$, $F_{UT,T}^{\sin(\phi-\phi_S)}$, $F_{UT}^{\sin(\phi+\phi_S)}$, $F_{UT}^{\sin(3\phi-\phi_S)}$, and $F_{LT}^{\cos(\phi-\phi_S)}$ are well-known. In position space, the expression for them can be found in ref.\cite{Boer:2011xd} and agrees with ours. It provides a cross-check for all the definitions and the computation code.

The NLP structure functions $F_{UU}^{\cos \phi}$, $F_{LU}^{\sin \phi}$, $F_{UL}^{\sin \phi}$, $F_{LL}^{\cos \phi}$, $F_{UT}^{\sin \phi_S}$, $F_{UT}^{\sin (2\phi-\phi_S)}$, $F_{LT}^{\cos \phi_S}$, and $F_{LT}^{\cos (2\phi-\phi_S)}$ can be compared with the literature only partially since the present formal derivation included previously overlooked effects and the complete NLO structure. We agree on all terms that could be compared with earlier computations (see below). There are two differences between our and earlier computations. They are
\begin{itemize}
\item The part of the genuine twist-three contribution proportional to $\mathbb{C}_I$.
\item The terms with derivatives of the Collins-Soper kernel.
\end{itemize}
These effects are critical for the consistency of the factorization theorem. Below we discuss these points in more detail. Both these effects were overlooked in the previous derivations \cite{Mulders:1995dh, Bacchetta:2006tn, Ebert:2021jhy} due to the assumption that partons carry strictly positive collinear momentum. This assumption is violated for genuine higher-twist contributions, which represent the quantum interference contribution. 

For the first time, we present the complete NLO structure for genuine terms, including evolution and coefficient functions. We emphasize that the consistent presentation of the factorization theorem requires the decomposition of twist-three distributions via quark-gluon-quark correlators since only this basis of twist-three operators is complete (see general discussion in ref.~\cite{Braun:2009mi, Braun:2011dg}). The traditional basis of bi-quark twist-three TMD distributions (e.g., \cite{Mulders:1995dh, Bacchetta:2006tn, Boer:2011xd}) is incomplete in the sense that NLO effects induce mixing with other operators (see explicit examples in sec.6 of ref.\cite{Rodini:2022wki}). The expression derived here is general, and its extension to higher perturbative orders requires only the computation of higher-order coefficient functions.

Eliminating the new terms, we can compare the LO part of our result with the expressions presented in the literature. The computation in ref.\cite{Bacchetta:2006tn} was performed and presented in momentum space and in terms of bi-quark TMD distributions. To make the comparison, we have used the relations between bi-quark TMD distributions and the (geometric) twist-three distributions derived in ref.\cite{Rodini:2022wki}. We agree with ref.\cite{Bacchetta:2006tn}. Also, we compare the kinematic corrections coming from the LP and NLP term with ref.\cite{Ebert:2021jhy}, where they are presented independently (see also appendix \ref{app:KPC}), and we agree with them as far as we can understand the notation. Let us note that this comparison is frame-dependent, and apparently, the frames used here and in ref.\cite{Ebert:2021jhy} are (potentially) different only at N$^2$LP. The kinematic part of the expression is also known as the Wandzura-Wilczek approximation; it has been studied in ref.\cite{Bastami:2018xqd} and also agrees with our result.

All TMD distributions at NLP obey the same evolution equation with respect to the rapidity parameter $\zeta$ (\ref{evol:rapidity}) with the same Collins-Soper kernel. It follows from the general consideration of rapidity divergences \cite{Vladimirov:2017ksc} or from the form of soft-function \cite{Ebert:2021jhy}. The same rapidity evolution is a fortunate fact that increases the universality of the TMD factorization approach and allows for the reuse of results of the precise extractions (such as \cite{Bacchetta:2022awv, Moos:2023yfa}) at NLP. Recently, the universality of evolution for twist-three TMD distributions has been tested with lattice computation \cite{Shu:2023cot}.

The field-mode decomposition is invariant under the variation of the rapidity separation of fields. On the level of cross-section, it required the invariance under the rescaling
\begin{eqnarray}
\zeta\to \alpha \zeta, \qquad \bar \zeta\to \frac{\zeta}{\alpha}.
\end{eqnarray}
The cross-section must be independent of the parameter $\alpha$, analogously to the independence of the parameter $\mu$. All terms without derivatives are trivially invariant under this transformation. 
Meanwhile, the terms involving the derivative with respect to $b$ are invariant only in combinations with the derivatives of the Collins-Soper kernel (\ref{def:mathring}). I.e., the combinations $\mathring{f}(\zeta)D(\bar \zeta)$ are boost-invariant, which is an important check of the result.

The evolution equations for the TMD distributions presented in sec.~\ref{sec:evol} are the consequence of operator structure. In that sense, they are independent entries to the factorization theorem, and all features of evolution should be independently realized within the coefficient function. In particular, the factorization theorem must provide the mixture mechanism between T-odd and T-even TMDPDFs of twist-three (\ref{def:evol-PDF}). This mechanism is realized by the terms proportional to $\mathbb{C}_I$. On the level of coefficient functions, the mixing terms are the ones that are presented in the second lines of (\ref{CR}) and (\ref{CI}). The confirmation of cancellation between logarithm parts and evolution equation for TMD distributions is done in ref.\cite{Vladimirov:2021hdn} and gives a strong check of our result. 

Generally speaking, all genuine NLP terms have a single hard coefficient function. However, it can be observed only in the position space; see ref.\cite{Vladimirov:2021hdn}. The transformation to the momentum space completely hides this universality because of the different support for momentum fractions in TMDPDFs and TMDFFs, producing different complex logarithms parts. This represents the primary reason why the coefficient function $\mathbb{C}_2$ (\ref{C2}) cannot be entirely obtained from $\mathbb{C}_R$ (\ref{CR}) (although the most part can be reconstructed by $x_i\to1/z_i$). Furthermore, the integrand is singular at $u_2=0$, and to define it completely, one needs to evaluate the pole contribution explicitly. It results in a novel contribution $\sim \mathbb{C}_I$ (\ref{CI}), already at LO. Therefore, practically, the genuine part of the NLP factorization theorem incorporates three different hard coefficient functions $\mathbb{C}_R$, $\mathbb{C}_I$ and $\mathbb{C}_2$, contrary to the assumption made in ref.\cite{Ebert:2021jhy}.

The LO term of the coefficient function $\mathbb{C}_I$ is $\delta(x_2)$ (\ref{CI}) and thus generates a leading contribution $\sim f_\bullet(-x,0,x)$. These terms are similar to the famous Qiu-Sterman mechanism in the collinear factorization of single-spin asymmetry \cite{Qiu:1991pp, Efremov:1981sh}. Such contributions are present in all structure functions via the $\mathcal{J}_n^{(I)}$-structures. 
Currently, there are no studies of such effects in TMD physics, so there is no criteria to estimate their size. Some of such contributions could be zero (at LO) due to symmetries or other effects. However, some contributions are definitely non-zero. Our estimation is based on the small-b computation performed in ref.\cite{Rodini:2022wki}. Using this computation, we can declare that functions $f_{\oplus T}(-x,0,x,b)$ and $h_\oplus(-x,0,x,b)$ are non-zero since they have a non-vanishing small-b limit. These distributions contribute to $F_{UU}^{\cos\phi}$ and $F_{UT}^{\sin \phi_S}$. The size of these contributions can be estimated by comparing LO small-b matching for $f_{\oplus T}$ and $h_\oplus$ with the small-b computations for twist-two distributions \cite{Scimemi:2018mmi, Rein:2022odl}. We estimate
\begin{eqnarray}
f_{\oplus T}(-x,0,x,b)\sim -\pi^{-1} f_{1T}^\perp (x,b),\qquad
h_\oplus(-x,0,x,b)\sim \pi^{-1} h_1^\perp(x,b).
\end{eqnarray}
So, the Qiu-Sterman-like contribution to the structure functions $F_{UU}^{\cos\phi}$, $F_{UT}^{\sin \phi_S}$ has approximately the same nonperturbative content as LP structure functions $F_{UU}^{\cos2\phi}$ and $F_{UT,T}^{\sin(\phi-\phi_S)}$. The main difference is the order of the Bessel function, which should numerically enhance the NLP terms. So,
\begin{eqnarray}\label{estimation}
F_{UU}^{\cos\phi}|_{\text{QS-like}} \gtrsim \frac{2M}{Q} F_{UU}^{\cos2\phi},
\qquad
F_{UT}^{\sin \phi_S}|_{\text{QS-like}} \gtrsim \frac{2M}{Q} F_{UT,T}^{\sin(\phi-\phi_S)}.
\end{eqnarray}
We conclude that the newly observed terms could produce  potentially large effects. Therefore, using the proper TMD factorization formula instead of a simplified one can lead to important phenomenological consequences already at LO. A more detailed investigation of practical implications will be done in future works.

It is instructive to investigate integrands' behavior in the small-b limit since some integrals are singular at first glance. Most TMD distributions have a regular form of small-b expansion, 
\begin{eqnarray}\label{small-b:reg}
f_{\text{reg}}(x,b;\mu,\zeta)\sim[C(\ln(\mu_{\text{OPE}}^2b^2;\mu,\zeta)\otimes F(\mu_{\text{OPE}})](x)+\mathcal{O}(b^2),
\end{eqnarray}
where $f$ is a TMD distribution, $F$ is a collinear distribution, $C$ is a coefficient function and $\otimes$ is an integral convolution in momentum fractions. A similar expression holds for TMDFFs. Note that $F$ can be twist-two or twist-three collinear distributions. All TMD distributions of twist-two have regular small-b expansion \cite{Moos:2020wvd}. Therefore, the derivatives of TMDPDFs of twist-two produce a singularity (we eliminate the double-logarithmic terms for simplicity)
\begin{eqnarray}
\mathring{f}_{\text{reg}}(x,b;\mu,\zeta)\sim \frac{a_s(\mu_{\text{OPE}})}{b^2}[C'(\ln(\mu_{\text{OPE}}^2b^2;\mu,\zeta)\otimes F(\mu_{\text{OPE}})](x)+\mathcal{O}(b^0).
\end{eqnarray}
At higher perturbative orders the singularity is stronger $\sim a_s^n \ln^{n-1}(b)/b^{2}$. The structure functions are finite because all singular terms are either accompanied by $b^2$ (see, e.g. (\ref{result:FUT:S})), or by a Bessel function of sufficiently high order (see, e.g. (\ref{result:FLL:1})). The structure functions $F_{UU}^{\cos\phi}$ and $F_{LL}^{\cos\phi}$ has the strongest behavior. Their integrands have logarithm singularity at $b\to0$, while integrands of all other structure functions vanish at this point.

Some of the TMD distributions of twist-three possess a more singular behavior
\begin{eqnarray}\label{small-b:sing}
f_{\text{sing}}(x,b;\mu,\zeta)\sim \frac{a_s(\mu_{\text{OPE}})}{M^2b^2}[C(\ln(\mu_{\text{OPE}}^2b^2;\mu,\zeta)\otimes F(\mu_{\text{OPE}})](x)+\mathcal{O}(b^0),
\end{eqnarray}
where $F$ is the collinear distribution of twist-two. In all cases, when the TMD distribution matches the collinear distribution with the lower twist, the mismatch in dimensions of operators is compensated by powers of $b$. The LO terms for small-b matching for singular distributions are presented in Appendix C of ref.\cite{Rodini:2022wki}. The singular distributions are
\begin{eqnarray}
f_{\text{sing}}\in \{D^\perp_{\ominus}+G_{\oplus}^\perp, f^\perp_{\ominus}-g_{\oplus}^\perp, ~f_{\oplus L}^\perp+g_{\ominus L}^\perp,~ 
h_{\ominus T}^{D\perp},~ h_{\ominus T}^{A\perp}\}.
\end{eqnarray}
Note that the distributions that mix with $f_\text{sing}$ by evolution are also singular but have the leading singularity NLO $\sim a_s^2/b^2$. Similarly to the kinematic term, we found that all singularities are compensated by either $b^2$ or by Bessel functions. Thus, all structure functions are well-defined.

\section{Conclusion}

We have presented the expression for the cross-section for the polarized SIDIS at next-to-leading power (NLP) and next-to-leading order (NLO) in TMD factorization. The main result of the paper is the collections of structure functions (\ref{result:FUUT} -- \ref{result:FLT:cos2S}). This work is the natural conclusion of the studies made in refs.\cite{Vladimirov:2021hdn, Rodini:2022wic, Rodini:2022wki}, where we elaborated the details of the TMD factorization at NLP.

Our computation non-trivially confirms the factorization hypothesis for SIDIS at NLP, which is not proven so-far but checked at NLO. In many aspects the NLP TMD factorization is similar to the LP one. Basically, there are only two additional elements -- the special rapidity divergences, and the complex part of the hard coefficient function. These elements appear during the transition from the factorized operator expression to the momentum representation of matrix elements. Therefore, we do not foresee any obstacles for the NLP factorization been valid at all perturbative orders, although the formal proof could be  essentially difficult.

The description of SIDIS at NLP involves TMD parton distribution functions (TMDPDFs) and TMD fragmentation functions (TMDFFs) of twist-two and twist-three. The twist-three TMD distributions are unexplored objects. The properties of the TMDPDFs of twist-three were considered in ref.\cite{Rodini:2022wki}, whereas the TMDFFs of twist-three are considered in this work for the first time, and represent one of the original parts of this work.

The derived expressions are largely consistent with previous computations, although some new terms have emerged that were previously overlooked. The main new findings are the terms with the quark-gluon-quark TMD correlators with the vanishing collinear part of gluon's momentum. These contributions are present in all NLP structure functions already at LO. The estimation of magnitude and practical importance of these terms is left for future works, but already now we can observe that some of them (\ref{estimation}) should be numerically sizable.

\addtocontents{toc}{\protect\setcounter{tocdepth}{1}}
\section*{Acknowledgments}
\addtocontents{toc}{\protect\setcounter{tocdepth}{2}}
We thank Ignazio Scimemi and Ian Balitsky for multiple discussions. We also thank Arturo Arroyo and Oscar del Rio for warning us about misprints in the original version of the paper. A.V. is funded by the \textit{Atracci\'on de Talento Investigador} program of the Comunidad de Madrid (Spain) No. 2020-T1/TIC-20204. A.V. is also supported by the Spanish Ministry grant PID2019-106080GB-C21. S.R. acknowledge the financial support from the physics department of Ecole Polytechnique.

\appendix

\section{Complex structure of genuine NLP term}
\label{app:complex-structure}

Each irreducible component of the hadronic tensor has a definite complexity, such that after convolution with the leptonic tensor the cross-section is real. However, individually elements of NLP factorization theorem are complex-valued. This is how the hadronic tensor is presented in the original derivation in ref.\cite{Vladimirov:2021hdn}. For practical purposes, one should rewrite this form explicitly revealing real and imaginary components. The computation is straightforward. However, the final expression is rather different from the starting one, and its origin is not transparent. Here we present some details of this computation.

This problem concerns only the genuine power correction term, since in LP and kinematic NLP terms all elements are real. In the notation of this work, the genuine NLP term computed in ref.\cite{Vladimirov:2021hdn} (explicitly, the second and the third lines of (6.17)) is
\begin{eqnarray}\label{app:W1}
&&\widetilde{W}_{\text{gNLP}}=i \sum_{n,m}
\Bigg\{
\\\nn &&\quad
\(\frac{\bar n^\mu }{q_-}-\frac{n^\mu}{q^+}\)\Tr[\gamma^{\rho}\Gamma_m^+ \gamma^\nu \Gamma_n^-]\Big[
\int [d\hat u]
C_2^\dagger\(\hat u_1,\hat u_2\)C_1
\mathbf{\Phi}_{\rho,21}^{[\Gamma^+_n]}(\hat u_{1,2,3};b)
\frac{\delta(x-\hat u_3)}{\hat u_2-i0}
\Delta_{11}^{[\Gamma_m^-]}(z,b)
\\\nn &&\qquad\qquad\qquad\qquad\qquad\qquad
-\int \frac{[d\hat w]}{\hat w_1\hat w_2}
C_2^\dagger\(\frac{1}{\hat w_1},\frac{1}{\hat w_2}\)C_1
\Phi_{11}^{[\Gamma^+_n]}(x,b)
\frac{\delta(z-\hat w_3)}{\hat w_2^{-1}+i0}
\mathbf{\Delta}_{\rho,21}^{[\Gamma_m^-]}(\hat w_{1,2,3};b)
\Big]
\\\nn &&
+
\(\frac{n^\mu}{q_+}-\frac{\bar n^\mu}{q^-}\)\Tr[\gamma^{\rho}\Gamma_n^- \gamma^\nu \Gamma_m^+]
\Big[
\int[d\hat u]
C_2^\dagger\(\hat u_1,\hat u_2\)C_1
\overline{\mathbf{\Phi}}_{\rho,21}^{[\Gamma_n^+]}(\hat u_{1,2,3};b)
\frac{\delta(x-\hat u_3)}{\hat u_2-i0}
\overline{\Delta}_{11}^{[\Gamma^-_m]}(z,b)
\\\nn &&\qquad\qquad\qquad\qquad\qquad\qquad
-
\int \frac{[d\hat w]}{\hat w_1\hat w_2}
C_2^\dagger\(\frac{1}{\hat w_1},\frac{1}{\hat w_2}\)C_1
\overline{\Phi}_{11}^{[\Gamma_n^+]}(x,b)
\frac{\delta(z-\hat w_3)}{\hat w_2^{-1}+i0}\overline{\mathbf{\Delta}}_{\rho,21}^{[\Gamma^-_m]}(\hat w_{1,2,3};b)
\Big]
\\\nn &&
+\(
\frac{n^\nu}{q_+}-\frac{\bar n^\nu}{q_-}\)\Tr[\gamma^\mu \Gamma_n^-\gamma^\rho \Gamma_m^+]
\Big[
\int [d\hat u]
C_1^\dagger C_2\(\hat u_3,\hat u_2\)
\overline{\mathbf{\Phi}}_{\rho,12}^{[\Gamma_n^+]}(\hat u_{1,2,3};b)
\frac{\delta(x+\hat u_1)}{\hat u_2-i0}
\overline{\Delta}_{11}^{[\Gamma_m^-]}(z,b)
\\\nn &&\qquad\qquad\qquad\qquad\qquad\qquad
-
\int \frac{[d\hat w]}{\hat w_2\hat w_3}
C_1^\dagger C_2\(\frac{1}{\hat w_3},\frac{1}{\hat w_2}\)
\overline{\Phi}_{11}^{[\Gamma_n^+]}(x,b)
\frac{\delta(z+\hat w_1)}{\hat w_2^{-1}+i0}
\overline{\mathbf{\Delta}}_{\rho,12}^{[\Gamma_m^-]}(\hat w_{1,2,3};b)
\Big]
\\\nn && 
+\(\frac{\bar n^\nu}{q_-}-\frac{n^\nu}{q_+}\)\Tr[\gamma^\mu \Gamma_m^+\gamma^\rho \Gamma_n^-]
\Big[
\int [d\hat u]
C_1^\dagger C_2\(\hat u_3,\hat u_2\)
\mathbf{\Phi}_{\rho,12}^{[\Gamma_n^+]}(\hat u_{1,2,3};b)
\frac{\delta(x+\hat u_1)}{\hat u_2-i0}
\Delta_{11}^{[\Gamma_m^-]}(z,b)
\\\nn &&\qquad\qquad\qquad\qquad\qquad\qquad
-
\int \frac{[d\hat w]}{\hat w_2\hat w_3}
C_1^\dagger C_2\(\frac{1}{\hat w_3},\frac{1}{\hat w_2}\)
\Phi_{11}^{[\Gamma_n^+]}(x,b)
\frac{\delta(z+\hat w_1)}{\hat w_2^{-1}+i0}
\mathbf{\Delta}_{\rho,12}^{[\Gamma_m^-]}(\hat w_{1,2,3};b)
\Big]\Bigg\},
\end{eqnarray}
where we refer to sec.\ref{sec:NLP-gen} and \ref{sec:TMDs} for the notation. In this expression, we explicitly keep the $\pm i0$ prescription for the gluon poles at $\hat u_2=0$ and $\hat w_2^{-1}=0$. The signs of $i0$ are related to the directions of gauge links of TMDPDFs and TMDFFs.

The coefficient function $C_1$ is the LP coefficient function (\ref{def:C1^2}). At NLO it reads
\begin{eqnarray}\label{def:C1}
C_1=1+a_sC_F\(-\mathbf{L}^2+3\mathbf{L}-8+\frac{\pi^2}{6}\)+\mathcal{O}(a_s^2),
\end{eqnarray}
where $C_F=(N_c^2-1)/2N_c^2$, $a_s=g^2/(4\pi)^2$, and
$$
\mathbf{L}=\ln\Big(\frac{-q^2-i0}{\mu^2}\Big).
$$
In the SIDIS kinematics the argument of logarithm is positive, and thus the coefficient function $C_1$ is real-valued, $C_1=C_1^\dagger$. The coefficient function $C_2$ is the coefficient function for quark-gluon-quark operator. At NLO it has been computed in ref.\cite{Vladimirov:2021hdn} (see eqn.(6.15)). In the SIDIS kinematics, it reads
\begin{eqnarray}\label{app:C2}
C_2(a,b)&=&1+a_s\Big[C_F\(-\mathbf{L}^2+\mathbf{L}-3+\frac{\pi^2}{6}\)
-C_A\frac{a+b}{a}\ln\(\frac{b}{a+b}-i0\) 
\\\nn &&
-\(C_F-\frac{C_A}{2}\)\frac{a+b}{b}\ln\(\frac{a}{a+b}-i0\) 
\Big(2\mathbf{L}+\ln\(\frac{a}{a+b}-i0\) -4\Big)\Big]+\mathcal{O}(a_s^2),
\end{eqnarray}
where $C_A=N_c$.  Note, that this formula is valid only for SIDIS ($q^2<0$). For processes with $q^2>0$ (Drell-Yan, or Semi-Inclusive annihilation) the complex part is more involved. 

Inspecting  the coefficient functions accompanying TMDPDFs of twist-three, we observe that $C_2^\dagger(\hat u_1,\hat u_2)$ has $\hat u_1+\hat u_2=-x<0$, while $C_2(\hat u_3,\hat u_2)$ has $\hat u_3+\hat u_2=x>0$, due to $\delta$-functions in (\ref{app:W1}). The individual signs  of $\hat u$'s are not defined, and thus logarithms could be complex-valued. We denote 
\begin{eqnarray}
\frac{C^\dagger(\hat u_1,\hat u_2)C_1}{u_2-i0}=\mathbb{C}_R(x,\hat u_2)+i \pi \mathbb{C}_I(x,\hat u_2),
\end{eqnarray}
where $x=-\hat u_1-\hat u_2$. Inserting the definition of $C_1$ and $C_2$, and resolving the complex structure we obtain
\begin{eqnarray}
&&\mathbb{C}_{R}(x,u_2)=
\frac{1}{(u_2)_+}+a_s\Bigg\{
2\frac{C_F}{(u_2)_+}\(-\mathbf{L}^2+2\mathbf{L}-\frac{11}{2}+\frac{\pi^2}{6}\)
\\\nn &&
\qquad
+2\(C_F-\frac{C_A}{2}\)\frac{1}{(u_2)_+}\frac{x}{u_2}\[\(\mathbf{L}-2+\frac{1}{2}\ln\(\frac{|x+u_2|}{x}\)\)\ln\(\frac{|x+u_2|}{x}\)
-\frac{\pi^2}{2}\theta(-x-u_2)\]
\\\nn &&
\qquad
+C_A\frac{x}{x+u_2}\[-\(\frac{\ln|u_2|}{u_2}\)_++\frac{\ln x}{(u_2)_+}
+\frac{\pi^2}{2}\delta(u_2)\]
\Bigg\}+\mathcal{O}(a_s^2)
\\
&&
\mathbb{C}_{I}(x,u_2)=
\delta(u_2)+a_s\Bigg\{
2C_F\[\delta(u_2)\(-\mathbf{L}^2+2\mathbf{L}-\frac{15}{2}+\frac{\pi^2}{6}\)\]
\\\nn &&
\qquad
+2\(C_F-\frac{C_A}{2}\)\[\delta(u_2)\mathbf{L}
+\frac{1}{(u_2)_+}\frac{x}{u_2}\(
\theta(-x-u_2)(\mathbf{L}-2)+\theta(-x-u_2)\ln\(\frac{|x+u_2|}{x}\)\)\]
\\\nn &&
\qquad
+C_A\[\delta(u_2)(\ln x+2)-\frac{\theta(u_2)}{(u_2)_+}\frac{x}{x+u_2}\]\Bigg\}
+\mathcal{O}(a_s^2).
\end{eqnarray}
Here, the ``plus''-distribution is defined as
\begin{eqnarray}
\int du \(f(u)\)_+g(u)=
\int du f(u)\(g(u)-g(0)\),
\end{eqnarray}
and arise from the decomposition
\begin{eqnarray}
\frac{1}{u-i0}=\frac{1}{(u)_+}-i\pi \delta(u),
\qquad
\frac{\ln(u-i0)}{u-i0}=\(\frac{\ln|u|}{u}\)_++\frac{\pi^2}{2} \delta(u)-i\pi \frac{\theta(-u)}{(u)_+}.
\end{eqnarray}

Next, we observe that the coefficient functions satisfy the relation
\begin{eqnarray}
C_1^\dagger C_2(\hat u_3,\hat u_2)=\(C_1 C^\dagger_2(-\hat u_3,-\hat u_2)\)^\dagger,
\end{eqnarray}
where we imposed constraints set by $\delta$-function for these terms. Thus, applying this relation to eqn.(\ref{app:W1}), and changing variables $\{\hat u_1,\hat u_2,\hat u_3\}\to \{-\hat u_3,-\hat u_2,-\hat u_1\}$ in these integrals, we find that coefficients of correspongin terms turn into
\begin{eqnarray}
\frac{C_1^\dagger C_2(\hat u_3,\hat u_2)}{\hat u_2+i0}
\to \frac{\(C_2(\hat u_1,\hat u_2)\)^\dagger}{-\hat u_2+i0}
=-\(\frac{C_2(\hat u_1,\hat u_2)}{\hat u_2+i0}\)^\dagger=-\mathbb{C}_R+i\pi \mathbb{C}_I.
\end{eqnarray}
In this way, we are able to express all coefficients in the terms of $\mathbb{C}_R$ and $\mathbb{C}_I$.

The similar procedure is done for the terms with twist-three TMDFF. Inspecting the coefficient functions accompanying TMDFF of twist-three, we observe that $C_2^\dagger(\hat w_1^{-1},\hat w_2^{-1})$ has $\hat w_1^{-1}+\hat w_2^{-1}=-z^{-1}<0$, and
$C_2(\hat w_3^{-1},\hat w_2^{-1})$ has $\hat w_3^{-1}+\hat w_2^{-1}=z^{-1}>0$. Analysis of complex structure gives
\begin{eqnarray}
C_1^\dagger C_2\(\frac{1}{\hat w_3},\frac{1}{\hat w_2}\)=C_2^\dagger\(\frac{-1}{\hat w_3},\frac{-1}{\hat w_2}\)C_1.
\end{eqnarray}
These coefficient functions are real-valued because the signs of $\hat w_i$ are fixed (due to the definition of the domain of TMDFF). We denote
\begin{eqnarray}
C^\dagger_2\(\frac{1}{\hat w_1},\frac{1}{\hat w_2}\)C_1=\frac{\mathbb{C}_2(z,\hat w_2)}{\hat w_2},
\end{eqnarray}
where $z=-\hat w_1\hat w_2/(\hat w_1+\hat w_2)$, and 
\begin{eqnarray}
\mathbb{C}_2(z,w_2)&=&
1+a_s\Big[
2C_F\(-\mathbf{L}^2+2\mathbf{L}-\frac{11}{2}+\frac{\pi^2}{6}\)
\\\nn &&
+2\(C_F-\frac{C_A}{2}\)\frac{w_2}{z}\(\mathbf{L}-2+\frac{1}{2}\ln\(1-\frac{z}{|w_2|}\)\)\ln\(1-\frac{z}{|w_2|}\)
\\\nn &&
-C_A\frac{|w_2|}{|w_2|-z}\ln\(\frac{z}{|w_2|}\)\Big]+\mathcal{O}(a_s^2).
\end{eqnarray}
Note, that variable $\hat w_2$ never reach the point $\hat w_2=0$, and does not receive the complex part from it. Performing same transformations as for terms with twist-three TMDPDFs, we find common a $\delta(z-\hat w_3)$, which set $\hat w_2<0$. It produces an extra sign-factor, due to $\hat w_2/|\hat w_2|=-1$.

Finally, we use the definitions (\ref{def:PHI-plus}, \ref{def:PHI-minus}) and replace
\begin{eqnarray}
\mathbf{\Phi}_{21}(\hat u_{1,2,3},b)&=&\mathbf{\Phi}_{\oplus}(\hat u_{1,2,3},b)-i \mathbf{\Phi}_{\ominus}(\hat u_{1,2,3},b),
\\
\mathbf{\Phi}_{21}(-\hat u_{3,2,1},b)&=&\mathbf{\Phi}_{\oplus}(\hat u_{1,2,3},b)+i \mathbf{\Phi}_{\ominus}(\hat u_{1,2,3},b),
\end{eqnarray}
and similar for $\overline{\mathbf{\Phi}}$, $\mathbf{\Delta}$ and $\overline{\mathbf{\Delta}}$. After minor simplifications we arrive at expression (\ref{def:WgNLP}).

\section{Evolution of twist-three TMD distributions at LO}
\label{app:evolution}

The evolution equation for the TMD distributions of twist-three were derived in ref.\cite{Vladimirov:2021hdn, Rodini:2022wki}. Some parts of these equations (namely, the elements of kernels $\mathbb{P}$) could be extracted from and compared with the results in the literature  \cite{Braun:2009mi, Braun:2009vc, Beneke:2017ztn}. The full description and derivation of the following equations is given in ref.\cite{Rodini:2022wki}. Here we present only the final expressions.

The evolution of twist-three TMD distributions with the respect to rapidity scales is given by LP equation
\begin{equation}
\zeta \frac{\partial}{\partial \zeta} F(b,\mu,\zeta) = -\mathcal{D}(b,\mu)F(b,\mu,\zeta),
\end{equation}
where $F$ is TMDPDF or TMDFF of twist-two or twist-three, and $\mathcal{D}$ is the Collins-Soper kernel \cite{Collins:1984kg}. The evolution with respect to the parameter $\mu$ is different for TMDPDFs and TMDFFs and is presented in the following sections.

\subsection{Evolution of TMDPDFs of twist-three}

The evolution equation for TMDPDFs of twist-three reads
\begin{eqnarray}\label{app:TMDPDF-evol}
\mu^2 \frac{d}{d\mu^2}\(\begin{array}{c}F_1\\ F_2\end{array}\) &=&
\(\frac{\Gamma_{\text{cusp}}}{2}\ln\(\frac{\mu^2}{\zeta}\)
+\Upsilon_{x_1x_2x_3}\)
\(\begin{array}{c}F_1\\ F_2\end{array}\)
\\\nn &&\qquad\qquad
+
\(\begin{array}{cc}
2\mathbb{P}_A &  2\pi \Theta_{x_1x_2x_3}\\
-2\pi\Theta_{x_1x_2x_3}     & 2\mathbb{P}_A
\end{array}\)\(\begin{array}{c}F_1\\ F_2\end{array}\),
\end{eqnarray}
where all distributions are functions of $(x_1,x_2,x_3,b;\mu,\zeta)$. The $\Gamma_{\text{cusp}}$ is the anomalous dimension of the light-like cusps of Wilson lines (see f.i.\cite{Moch:2004pa}). The LO expression for the function $\Upsilon_{x_1x_2x_3}$ is
\begin{eqnarray}\label{app:upsilon}
\Upsilon_{x_1x_2x_3}=a_s\[3C_F+C_A\ln\(\frac{|x_3|}{|x_2|}\)+2\(C_F-\frac{C_A}{2}\)\ln\(\frac{|x_3|}{|x_1|}\)\]+\mathcal{O}(a_s^2).
\end{eqnarray}
The function $\Theta$ is discontinuous, and at LO it reads
\begin{equation}
\Theta_{x_1x_2x_3} = a_s \times \begin{cases}
\frac{C_A}{2} & x_{1,2,3} \in (+--)\\
-C_F+\frac{C_A}{2} & x_{1,2,3} \in (+-+)\\
0 & x_{1,2,3} \in (--+)\\
-\frac{C_A}{2} & x_{1,2,3} \in (-++)\\
C_F-\frac{C_A}{2} & x_{1,2,3} \in (-+-)\\
0 & x_{1,2,3} \in (++-)\\
\end{cases}\qquad +\mathcal{O}(a_s^2)
\end{equation}
where $x_{1,2,3}\in (+--)$ means the region $(x_1>0, x_2<0, x_3<0)$, and similarly for all the others. Importantly, the equation (\ref{app:TMDPDF-evol}) is written for the SIDIS-like TMDPDFs which have Wilson lines pointing to $+\infty$. For the case of DY-like TMDPDFs with Wilson lines pointing to $-\infty$, the mixing terms $\Theta$ should be taken with opposite signs. The kernel $\mathbb{P}_A$ is the integral kernel that acts on the TMDPDF
\begin{align}\label{def:PA-PDF}
\mathbb{P}^A_{x_1x_2}&\Phi_{21}(x_1,x_2,x_3) = -\frac{a_s}{2} \Bigg\{ C_A \delta_{x_2,0} \Phi_{21}(x_1,0,x_3) + C_A \int_{-\infty}^\infty dv \\
& \Bigg[ \ta (x_2+v)\Phi_{21}(x_1,x_2,x_3)-x_2\Phi_{21}(x_1-v,x_2+v,x_3)\tc \frac{x_2\Theta(x_2,v)}{v(x_2+v)^2} \nn\\
& -\ta \Phi_{21}(x_1,x_2,x_3)-\Phi_{21}(x_1-v,x_2+v,x_3)\tc \frac{x_1 \Theta(x_1,-v)}{v(x_1-v)}\Bigg] \nn\\
&- C_A \int_{-\infty}^\infty dv \frac{\Phi_{21}(x_1-v,x_2+v,x_3)}{x_3^2} \Bigg[\frac{x_2^2(v+2x_2+x_1)}{(x_2+v)^2}\Theta(x_2,v)+\frac{x_1(2x_2+x_1)}{x_1-v}\Theta(x_1,-v) \Bigg] \nn\\\nn
&+ 2\ta C_F-\frac{C_A}{2}\tc \int_{-\infty}^\infty dv \frac{\Phi_{21}(x_2+v,x_1-v,x_3)}{x_3^2} \Bigg[ \frac{x_2^2}{x_2+v}\Theta(x_2,v)
\\\nn &+\frac{x_1(x_2x_1-2vx_2-vx_1)}{(x_1-v)^2}\Theta(x_1,-v)\Bigg]
\Bigg\}+\mathcal{O}(a_s^2),
\end{align}
where $\Theta(a,b) = \theta(a)\theta(b)-\theta(-a)\theta(-b)$.

Importantly, the equation (\ref{app:TMDPDF-evol}) is valid only for the following pairs of TMD distributions of twist-three
\begin{eqnarray}
\(\begin{array}{c}F_1\\ F_2\end{array}\)_A \in \Bigg\{
\(\begin{array}{c}f_{\oplus T}+g_{\ominus T}\\ f_{\ominus T}-g_{\oplus T}\end{array}\),
\(\begin{array}{c}f_{\oplus}^\perp+g_{\ominus}^\perp\\ f_{\ominus}^\perp-g_{\oplus}^\perp\end{array}\),
\(\begin{array}{c}f_{\oplus L}^\perp+g_{\ominus L}^\perp\\ f_{\ominus L}^\perp-g_{\oplus L}^\perp\end{array}\),
\(\begin{array}{c}f_{\oplus T}^\perp+g_{\ominus T}^\perp\\ f_{\ominus T}^\perp-g_{\oplus T}^\perp\end{array}\),
\\\nn
\(\begin{array}{c}h_\oplus\\ h_\ominus\end{array}\),
\(\begin{array}{c}h_{\oplus L}\\ h_{\ominus L}\end{array}\),
\(\begin{array}{c}h_{\oplus T}^{D\perp}\\ h_{\ominus T}^{D\perp}\end{array}\),
\(\begin{array}{c}h_{\oplus T}^{A\perp}\\ h_{\ominus T}^{A\perp}\end{array}\)\Bigg\}.
\end{eqnarray}
For the remaining distributions are
\begin{eqnarray}\label{app:TMD_B}
\(\begin{array}{c}F_1\\ F_2\end{array}\)_B \in \Bigg\{
\(\begin{array}{c}f_{\oplus T}-g_{\ominus T}\\ f_{\ominus T}+g_{\oplus T}\end{array}\),
\(\begin{array}{c}f_{\oplus}^\perp-g_{\ominus}^\perp\\ f_{\ominus}^\perp+g_{\oplus}^\perp\end{array}\),
\(\begin{array}{c}f_{\oplus L}^\perp-g_{\ominus L}^\perp\\ f_{\ominus L}^\perp+g_{\oplus L}^\perp\end{array}\),
\(\begin{array}{c}f_{\oplus T}^\perp-g_{\ominus T}^\perp\\ f_{\ominus T}^\perp+g_{\oplus T}^\perp\end{array}\),
\\\nn
\(\begin{array}{c}h_\oplus^\perp\\ h_\ominus^\perp\end{array}\),
\(\begin{array}{c}h_{\oplus L}^\perp\\ h_{\ominus L}^\perp\end{array}\),
\(\begin{array}{c}h_{\oplus T}^{S\perp}\\ h_{\ominus T}^{S\perp}\end{array}\),
\(\begin{array}{c}h_{\oplus T}^{T\perp}\\ h_{\ominus T}^{T\perp}\end{array}\)\Bigg\}.
\end{eqnarray}
Their evolution equation is same as (\ref{def:PA-PDF}) but with $\mathbb{P}_A$ replaced by $\mathbb{P}_B$. The later is given in eqn.(3.16) of ref.\cite{Rodini:2022wki}. The distributions (\ref{app:TMD_B}) do not appears in the NLP factorization theorem for SIDIS or Drell-Yan.

\subsection{Evolution of TMDFFs of twist-three}
\label{app:TMDFF-evol}

In position space, the evolution equation for TMDFF of twist-three has identically the same form as for TMDPDF. However, the transformation to the momentum-fraction space makes it look different, since the support regions for TMDPDFs and TMDFFs are different. In particular, in the case of TMDFF the evolution preserves the sign of momentum-fractions $z$'s, and thus does not generate complex parts. As a result, there is no mixing terms, and also the $\theta$-function structure simplifies.

The evolution equation reads
\begin{eqnarray}\label{eq:TMDFF-evol}
\mu^2 \frac{d}{d\mu^2}\Delta_A=
\Big[
\frac{\Gamma_{\text{cusp}}}{2}\ln\(\frac{\mu^2}{\zeta}\)+\Upsilon_{z_1z_2z_3}^{\text{FF}}+\mathbb{P}_A^{\text{FF}}\Big]\Delta_A,
\end{eqnarray}
where TMDFFs are functions of $(z_1,z_2,z_3,b;\mu,\zeta)$. The $\Gamma_{\text{cusp}}$ is the anomalous dimension of the light-like cusps of Wilson lines. The LO expression for the function $\Upsilon^{\text{FF}}_{z_1z_2z_3}$ is
\begin{eqnarray}
\Upsilon^{\text{FF}}_{z_1z_2z_3}=a_s\[3C_F+C_A\ln\(\frac{|z_2|}{|z_3|}\)+2\(C_F-\frac{C_A}{2}\)\ln\(\frac{|z_1|}{|z_3|}\)\]+\mathcal{O}(a_s^2).
\end{eqnarray}
Note, that $\Upsilon^{\text{FF}}_{z_1z_2z_3}=\Upsilon_{x_1x_2x_3}(x_i\to z_i^{-1})$ (\ref{app:upsilon}). The action of the integral kernel $\mathbb{P}^{\text{FF}}_A$ to TMDFF reads 
\begin{eqnarray}\label{app:PA-FF}
&&\mathbb{P}^{\text{FF}}_A\Delta(z_1,z_2,z_3) =
-\frac{a_s}{2} \Bigg\{ C_A \int_{-\infty}^\infty \frac{dv}{v}  \Bigg[
- \ta \Delta(z_1,z_2,z_3)-\frac{\Delta\ta \frac{z_1}{1-vz_1}, \frac{z_2}{1+vz_2},z_3\tc}{(1-v z_1)(1+vz_2)^2}\tc\frac{\theta(-v)}{1+vz_2} \nn\\
&&
\qquad\qquad\qquad
+ \ta \Delta(z_1,z_2,z_3)-\frac{\Delta\ta \frac{z_1}{1-vz_1}, \frac{z_2}{1+vz_2},z_3\tc}{(1-v z_1)(1+vz_2)}\tc\frac{\theta(v)}{1-vz_1}
\Bigg] \\
&& +
C_A \int_{-\infty}^\infty dv \frac{\Delta\ta \frac{z_1}{1-vz_1}, \frac{z_2}{1+vz_2},z_3\tc}{(1-v z_1)(1+vz_2)}\frac{z_1z_2}{(z_1+z_2)^2} 
\Bigg( \frac{2z_1+z_2+vz_1z_2}{(1+vz_2)^2}\theta(-v) + \frac{2z_1+z_2}{1-vz_1}\theta(v)\Bigg) \nn \\
&&\nn
-2\ta C_F -\frac{C_A}{2}\tc \int_{-\infty}^\infty dv \frac{\Delta\ta \frac{z_2}{1+vz_2}, \frac{z_1}{1-vz_1},z_3\tc}{(1-v z_1)(1+vz_2)}\frac{z_1^2z_2}{(z_1+z_2)^2}
\Bigg(\frac{\theta(-v)}{1+vz_2}+\frac{1-2vz_1-vz_2}{(1-vz_1)^2}\Theta(v)\Bigg)
\Bigg\} ,
\end{eqnarray}
where $\theta$ is the Heaviside theta-function. This equation was obtained from the universal position-space expression by computing the Fourier transform (\ref{def:DELTA21}). It can be compared with the kernel for TMDPDF (\ref{def:PA-PDF}) by replacing $\Phi(a,b,c)\to acb\Delta(a^{-1},b^{-1},c^{-1})$, and subsequent replacement $x_i\to z_i^{-1}$, and multiplication by global factor $|z_1z_2z_3|^{-1}$. Also, in derivation of (\ref{app:PA-FF}) we took into account that $\Delta\neq0$ only in the region $z_{1,2}<0$ and $z_3>0$.

The evolution equation (\ref{eq:TMDFF-evol}) is valid only for the following TMDFFs
\begin{eqnarray}
\Delta_A\in \{
D_{\oplus}^\perp-G_{\ominus}^\perp, D_{\ominus}^\perp+G_{\oplus}^\perp,
H_\oplus, H_\ominus\}.
\end{eqnarray}
For other TMDFFs
\begin{eqnarray}\label{app:DELTAB}
\Delta_B\in \{
D_{\oplus}^\perp+G_{\ominus}^\perp, D_{\ominus}^\perp-G_{\oplus}^\perp,
H_\oplus^\perp, H_\ominus^\perp\},
\end{eqnarray}
the kernel $\mathbb{P}_A$ must be replaced by the kernel $\mathbb{P}_B$ which is
\begin{eqnarray}\label{app:PB-FF}
\mathbb{P}^{\text{FF}}_B\Delta(z_1,z_2,z_3)& =& -\frac{a_s}{2} \Bigg\{
C_A \int_{-\infty}^\infty \frac{dv}{v} 
\Bigg[ -\ta \Delta(z_1,z_2,z_3)-\frac{\Delta\ta \frac{z_1}{1-vz_1}, \frac{z_2}{1+vz_2},z_3\tc}{(1-v z_1)(1+vz_2)^2}\tc\frac{\theta(-v)}{1+vz_2} \nn
\\&&\nn
+ \ta \Delta(z_1,z_2,z_3)-\frac{\Delta\ta \frac{z_1}{1-vz_1}, \frac{z_2}{1+vz_2},z_3\tc}{(1-v z_1)(1+vz_2)}\tc\frac{\theta(-v)}{1-vz_1}
\Bigg]\nn \\
&& +2\ta C_F-\frac{C_A}{2}\tc \int_{-\infty}^\infty dv \frac{\Delta\ta \frac{z_2}{1+vz_2}, \frac{z_1}{1-vz_1},z_3\tc}{(1-v z_1)(1+vz_2)}\frac{z_1}{(1-vz_1)^2}\Theta(v)
\Bigg\}.\nn    
\end{eqnarray}
The kernel $\mathbb{P}_B$ has been derived using the same technique as the kernel $\mathbb{P}_A$ (\ref{app:PA-FF}). To our best knowledge, the evolution equations for TMDFFs of twist-three are discussed here for the first time.

\section{Kinematic power corrections separately}
\label{app:KPC}

The kinematic power corrections have two sources. The first source is the expansion of kinematic variables from the contraction of LP hadronic tensor, for shortness we call it LP$'$. The second source is the NLP hadronic tensor, for shortness we call it kNLP. In this appendix we present the expressions for both terms separately, and explain the relation which help to reduce them to the simple form.

Expanding the contraction of $L_{\mu\nu}W^{\mu\nu}_{\text{LP}}$ up to NLP we obtain the LP$'$ contribution (here only the terms which produce non-zero NLP contribution)
\begin{eqnarray}
F_{UU}^{\cos \phi}|_{\text{LP}'}&=&-\frac{|p_\perp|}{zQ}\(
\mathcal{J}_0[f_1D_1]+\mathcal{J}_2[h_1^\perp H_1^\perp]\),
\\
F_{UL}^{\sin \phi}|_{\text{LP}'}&=&-\frac{|p_\perp|}{zQ}
\mathcal{J}_2[h_{1L}H_1^\perp],
\\
F_{LL}^{\cos\phi}|_{\text{LP}'}&=&-\frac{|p_\perp|}{zQ}
\mathcal{J}_0[g_{1}D_1],
\\
F_{UT}^{\sin \phi_S}|_{\text{LP}'}&=&-\frac{|p_\perp|}{zQ}
\mathcal{J}_1\[\frac{1}{2}f_{1T}^\perp D_1+h_{1} H_1^\perp\],
\\
F_{UT}^{\sin(2\phi-\phi_S)}|_{\text{LP}'}&=&-\frac{|p_\perp|}{zQ}
\(-\frac{1}{2}\mathcal{J}_1[f_{1T}^\perp D_1]+\frac{1}{4}\mathcal{J}_3[h_{1T}^\perp H_1^\perp]\),
\\
F_{LT}^{\cos\phi_S}|_{\text{LP}'}&=&-\frac{|p_\perp|}{zQ}\frac{1}{2}\mathcal{J}_1[g_{1T}^\perp D_1],
\\
F_{LT}^{\cos(2\phi-\phi_S)}|_{\text{LP}'}&=&-\frac{|p_\perp|}{zQ}\frac{1}{2}\mathcal{J}_1[g_{1T}^\perp D_1],
\end{eqnarray}
where $\mathcal{J}_n$ is defined in eqn.(\ref{def:JN}). These expression agree with the blue-font part (denoted as kinematic corrections) of the expressions presented in sec.5.2.4 of ref.\cite{Ebert:2021jhy}.

The kNLP terms read
\begin{eqnarray}\label{app:11}
F_{UU}^{\cos \phi}|_{\text{kNLP}}&=&\frac{M}{Q}\mathcal{J}_1\[
\mathring{f}_1D_1-f_1\mathring{D}_1
-M^2 |b|^2 (\mathring{h}_1^\perp H_1^\perp-h_1^\perp \mathring{H}_1^\perp)
\],
\\
F_{UL}^{\sin \phi}|_{\text{kNLP}}&=&\frac{M}{Q}\mathcal{J}_1\[
-M^2 |b|^2 (\mathring{h}^\perp_{1L} H_1^\perp- h^\perp_{1L} \mathring{H}_1^\perp)
\],
\\
F_{LL}^{\cos\phi}|_{\text{kNLP}}&=&\frac{M}{Q}
\mathcal{J}_1\[\mathring{g}_1D_1-g_1\mathring{D}_1\],
\\
F_{UT}^{\sin \phi_S}|_{\text{kNLP}}&=&\frac{M}{Q}
\mathcal{J}_0\Big[-f_{1T}^\perp D_1+2h_{1} H_1^\perp
+\frac{b^2M^2}{2}\(f_{1T}^\perp \mathring{D}_1-\mathring{f}_{1T}^\perp D_1\)
\\\nn &&
\qquad
+b^2M^2\(h_{1} \mathring{H}_1^\perp-\mathring{h}_{1} H_1^\perp\)
\Big],
\\
F_{UT}^{\sin(2\phi-\phi_S)}|_{\text{kNLP}}&=&\frac{M}{Q}\mathcal{J}_2\Big[
-\frac{1}{2}h_{1T}^\perp H_1^\perp
+\frac{1}{2}\(f_{1T}^\perp \mathring{D}_1-\mathring{f}_{1T}^\perp D_1\)
\\\nn &&
\qquad
+\frac{M^2 |b|^2}{4} \(h_{1T}^\perp \mathring{H}_1^\perp-\mathring{h}^\perp_{1T} H_1^\perp\)
\Big],
\\
F_{LT}^{\cos\phi_S}|_{\text{kNLP}}&=&\frac{M}{Q}\mathcal{J}_0\[
-g_{1T}^\perp D_1
+\frac{b^2M^2}{2}\(g_{1T}^\perp\mathring{D}_1-\mathring{g}_{1T}D_1\)
\],
\\\label{app:22}
F_{LT}^{\cos(2\phi-\phi_S)}|_{\text{kNLP}}&=&\frac{M}{2Q}\mathcal{J}_2\[
\mathring{g}_{1T}D_1-g_{1T}^\perp\mathring{D}_1
\],
\end{eqnarray}
where the definition of operation $\mathring{A}$ is
\begin{eqnarray}\nn
\mathring{f}(x,b;\mu,\zeta)&=&\frac{2}{M^2}\[\frac{\partial }{\partial |b|^2}+\frac{1}{2}\ln\(\frac{\zeta}{\bar \zeta}\)\(\frac{\partial \mathcal{D}(b,\mu)}{\partial |b|^2}\)\]f(x,b;\mu,\zeta),
\\\nn
\mathring{D}(x,b;\mu,\bar \zeta)&=&\frac{2}{M^2}\[\frac{\partial }{\partial |b|^2}-\frac{1}{2}\ln\(\frac{\zeta}{\bar \zeta}\)\(\frac{\partial \mathcal{D}(b,\mu)}{\partial |b|^2}\)\]D(x,b;\mu,\bar \zeta).
\end{eqnarray}
Here, the first equation is for TMDPDF, and the second is for TMDFF. The expressions (\ref{app:11}-\ref{app:22}) agree with the green-font part (denoted as $\mathcal{P}_\perp$ corrections) of expressions presented in sec.5.2.4 of ref.\cite{Ebert:2021jhy}, to our best understanding.

It is important to note that these corrections are frame-dependent. A small rotation of a vector $n^\mu$ (or $\bar n^\mu$) does not alter the counting for the field mode (if the transformation preserves $n^2=\bar n^2=0$), and thus preserves the structure of factorization theorem \cite{Manohar:2002fd}. However, the some of LP$'$ and kNLP terms is frame-invariant up to N$^2$LP. We have explicitly confirmed this statement for $F_{UU}^{\cos\phi}$ term.

The terms $\mathcal{J}_n(f_1\mathring{f_2})$ can be rewritten via the terms $\mathcal{J}_n(\mathring{f_1}f_2)$ using the integration by parts. These relations are
\begin{eqnarray}
\mathcal{J}_n(f_1\mathring{f_2})&=&
-\frac{|p_\perp|}{zM}\mathcal{J}_{n-1}(f_1f_2)-\mathcal{J}_{n}(\mathring{f_1}f_2),\quad n>0,
\\
\mathcal{J}_n(|b|^2M^2 f_1\mathring{f_2})&=&
\frac{|p_\perp|}{zM}\mathcal{J}_{n+1}(f_1f_2)-
2(n+1)\mathcal{J}_{n}(f_1f_2)
-\mathcal{J}_{n}(|b|^2M^2\mathring{f_1}f_2),\quad n\geqslant 0,
\end{eqnarray}
where $f_1$ and $f_2$ are functions of $|b|^2$. Applying these relations one can eliminate the contributions of LP$'$-terms. We found that for all terms this can be achieved by turning the derivative from TMDFF to TMDPDF. This operation also eliminates some non-derivative kNLP terms. The result is presented in eqn.(\ref{result:FUUT}-\ref{result:FLT:cos2S}).

\bibliographystyle{bibliostyle}

\providecommand{\href}[2]{#2}\begingroup\raggedright\endgroup

\end{document}